\documentclass[aps,prb,twocolumn,footinbib,longbibliography]{revtex4}
\usepackage{amsmath,amsthm}
\usepackage{cancel}
\usepackage{amsfonts}
\usepackage[utf8]{inputenc}
\usepackage{graphicx}
\usepackage{xcolor}
\usepackage{bm}

\def\nbC{{\mathchoice {\setbox0=\hbox{$\displaystyle\rm C$}%
\hbox{\hbox to0pt{\kern0.4\wd0\vrule height0.9\ht0\hss}\box0}}
{\setbox0=\hbox{$\textstyle\rm C$}\hbox{\hbox
to0pt{\kern0.4\wd0\vrule height0.9\ht0\hss}\box0}}
{\setbox0=\hbox{$\scriptstyle\rm C$}\hbox{\hbox
to0pt{\kern0.4\wd0\vrule height0.9\ht0\hss}\box0}}
{\setbox0=\hbox{$\scriptscriptstyle\rm C$}\hbox{\hbox
to0pt{\kern0.4\wd0\vrule height0.9\ht0\hss}\box0}}}}
\def\nbQ{{\mathchoice {\setbox0=\hbox{$\displaystyle\rm
Q$}\hbox{\raise
0.15\ht0\hbox to0pt{\kern0.4\wd0\vrule height0.8\ht0\hss}\box0}}
{\setbox0=\hbox{$\textstyle\rm Q$}\hbox{\raise
0.15\ht0\hbox to0pt{\kern0.4\wd0\vrule height0.8\ht0\hss}\box0}}
{\setbox0=\hbox{$\scriptstyle\rm Q$}\hbox{\raise
0.15\ht0\hbox to0pt{\kern0.4\wd0\vrule height0.7\ht0\hss}\box0}}
{\setbox0=\hbox{$\scriptscriptstyle\rm Q$}\hbox{\raise
0.15\ht0\hbox to0pt{\kern0.4\wd0\vrule height0.7\ht0\hss}\box0}}}}
\def\nbT{{\mathchoice {\setbox0=\hbox{$\displaystyle\rm
T$}\hbox{\hbox to0pt{\kern0.3\wd0\vrule height0.9\ht0\hss}\box0}}
{\setbox0=\hbox{$\textstyle\rm T$}\hbox{\hbox
to0pt{\kern0.3\wd0\vrule height0.9\ht0\hss}\box0}}
{\setbox0=\hbox{$\scriptstyle\rm T$}\hbox{\hbox
to0pt{\kern0.3\wd0\vrule height0.9\ht0\hss}\box0}}
{\setbox0=\hbox{$\scriptscriptstyle\rm T$}\hbox{\hbox
to0pt{\kern0.3\wd0\vrule height0.9\ht0\hss}\box0}}}}
\def\nbS{{\mathchoice
{\setbox0=\hbox{$\displaystyle     \rm S$}\hbox{\raise0.5\ht0%
\hbox to0pt{\kern0.35\wd0\vrule height0.45\ht0\hss}\hbox
to0pt{\kern0.55\wd0\vrule height0.5\ht0\hss}\box0}}
{\setbox0=\hbox{$\textstyle        \rm S$}\hbox{\raise0.5\ht0%
\hbox to0pt{\kern0.35\wd0\vrule height0.45\ht0\hss}\hbox
to0pt{\kern0.55\wd0\vrule height0.5\ht0\hss}\box0}}
{\setbox0=\hbox{$\scriptstyle      \rm S$}\hbox{\raise0.5\ht0%
\hboxto0pt{\kern0.35\wd0\vrule height0.45\ht0\hss}\raise0.05\ht0%
\hbox to0pt{\kern0.5\wd0\vrule height0.45\ht0\hss}\box0}}
{\setbox0=\hbox{$\scriptscriptstyle\rm S$}\hbox{\raise0.5\ht0%
\hboxto0pt{\kern0.4\wd0\vrule height0.45\ht0\hss}\raise0.05\ht0%
\hbox to0pt{\kern0.55\wd0\vrule height0.45\ht0\hss}\box0}}}}
\def\nbZ{{\mathchoice {\hbox{$\sf\textstyle Z\kern-0.4em Z$}}
{\hbox{$\sf\textstyle Z\kern-0.4em Z$}}
{\hbox{$\sf\scriptstyle Z\kern-0.3em Z$}}
{\hbox{$\sf\scriptscriptstyle Z\kern-0.2em Z$}}}}
\begin{document}

\title{Criticality  of the random field Ising model in and out of equilibrium: a nonperturbative functional renormalization group description}

\author{Ivan Balog} \email{balog@ifs.hr}
\affiliation{Institute of Physics, P.O.Box 304, Bijeni\v{c}ka cesta 46, HR-10001 Zagreb, Croatia}

\author{Gilles Tarjus} \email{tarjus@lptmc.jussieu.fr}
\affiliation{LPTMC, CNRS-UMR 7600, Universit\'e Pierre et Marie Curie,
bo\^ite 121, 4 Pl. Jussieu, 75252 Paris c\'edex 05, France}

\author{Matthieu Tissier} \email{tissier@lptmc.jussieu.fr}
\affiliation{LPTMC, CNRS-UMR 7600, Universit\'e Pierre et Marie Curie,
bo\^ite 121, 4 Pl. Jussieu, 75252 Paris c\'edex 05, France}

\date{\today}

\begin{abstract}

We show that, contrary to previous suggestions based on computer simulations or erroneous theoretical treatments, the critical points of the random-field Ising model out of equilibrium, when quasi-statically changing the applied source at zero temperature, and in equilibrium are not in the same universality class below some critical dimension $d_{DR}\approx 5.1$. We demonstrate this by implementing a non-perturbative functional renormalization group for the associated dynamical field theory. Above $d_{DR}$, the avalanches, which characterize the evolution of the system at zero temperature, become irrelevant at large distance, and hysteresis and equilibrium critical points are then controlled by the same fixed point. We explain how to use computer simulation and finite-size scaling to check the correspondence between in and out of equilibrium criticality in a far less ambiguous way than done so far.

\end{abstract}


\maketitle

\section{Introduction}
\label{introduction}

The random-field Ising model is one of the paradigmatic models for investigating collective behavior resulting from the interplay of interactions and quenched disorder. The system can be studied either in equilibrium, where it is known to display a paramagnetic to ferromagnetic phase transition with a critical point in dimension $d>2$,\cite{imry-ma75,bricmont,nattermann98} and out of equilibrium, when driven at zero temperature under the influence of a slowly changed external magnetic field. The latter situation gives rise to a hysteresis curve in the magnetization versus magnetic field plane and, in the quasi-static limit of an infinitely slowly increased or decreased applied field, to a critical point on both the ascending (increasing field) and the descending (decreasing field) branches of the hysteresis loop.\cite{sethna93,dahmen96,perkovic,sethna05} These out-of-equilibrium critical points take place at nonzero values of the applied magnetic field and separate a high-disorder phase characterized by a continuous hysteresis curve from a low-disorder phase with a discontinuous hysteresis curve. 

Quite strikingly, despite the fact that one critical point is at equilibrium and at zero external field while the others are out-of-equilibrium and at nonzero values of the applied external field, and that they take place at different values of the disorder strength (see the sketch in Fig. 1), the two types of criticality are characterized by critical exponents that have been found very close in numerical simulations.\cite{maritan94,perez04,liu09}

\begin{figure}[ht]
\begin{center}
\includegraphics[width=\linewidth]{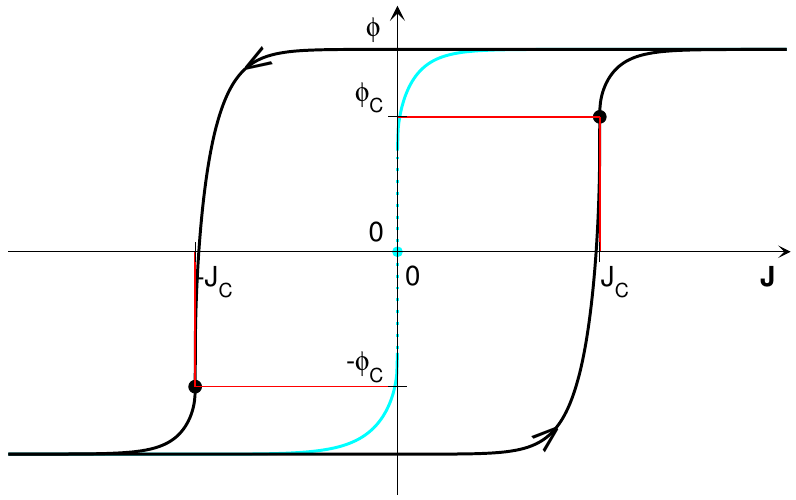}
\caption{RFIM at zero temperature: Schematic illustration of the hysteresis loop (black) and of the equilibrium curve (blue) in the magnetization ($m$ or $\phi$) versus applied magnetic field ($H$ or $J$) representation. For the chosen value of the disorder strength, the system has two (symmetric) out-of-equilibrium critical points (black dots) on the ascending and the descending branches, but a first-order transition in equilibrium (the equilibrium critical random-field strength is larger than the out-of-equilibrium one\cite{dahmen96,perkovic,perez04}).}
\end{center}
\end{figure}

A further key point of interest concerns the presence of  ``avalanches'' and their characteristics in and out of equilibrium. Abrupt changes in the system's configuration indeed generically take place at zero temperature in the presence of quenched disorder. When the system is slowly (quasi-statically) driven, these changes take the form of avalanches  that  may span a broad range of sizes and in the case of the RFIM become scale-free at criticality. The signature of these events is a ``crackling noise''\cite{sethna01} which can be found in a great variety of situations.\cite{sethna01,fisher98,dahmen-zion,bertotti,bouchaud13,hallock96,planes,detcheverry04,nematic2}  How does this transpose to the RFIM in equilibrium? First, one should keep in mind that the critical behavior of the RFIM at any temperature is controlled in a renormalization-group sense by a ``zero-temperature fixed point''.\cite{villain84,fisher86} One can then access the generic equilibrium critical behavior by directly studying the model at zero temperature. Second, although, strictly speaking, the RFIM in equilibrium is not driven, it goes at zero temperature from one ground state to another as a function of the external  magnetic field.\cite{frontera00,dukowski03,wu05,liu07} Discontinuous changes of the ground state represent ``static avalanches'' or ``shocks''\cite{BBM} (but there is no actual dynamics associated with them). Again, it has been found in $3$-$d$ computer simulations that static and dynamic avalanches are characterized by integrated size-distribution and correlation functions that appear indistinguishable within numerical accuracy.\cite{liu09}

The question we address in this work is that of the connection between in and out of equilibrium criticality in the RFIM. The theoretical framework for a proper resolution of this question is the renormalization group (RG): The critical behaviors are the same, and are therefore in the same universality class, provided  they are controlled by the same RG fixed point.  In the RG context, the signature of the avalanches appears as a nonanalytic functional dependence of the renormalized cumulants of the disorder, in the form of a linear ``cusp'' in the field dependence of the second- and higher-order cumulants of the renormalized random field or force. The relevant RG theory must therefore be functional.\cite{fisher86b,nattermann,narayan92,feldman,FRGledoussal-chauve,FRGledoussal-wiese,tarjus04,tissier06,tissier11}

The similarity of the in and out of equilibrium criticalities in the RFIM appears surprising in light of what has been found for a related but simpler disordered model, an elastic manifold in a random environment. The latter model has a scale-free behavior both in equilibrium, where the pinned manifold is in a rough phase, and out of equilibrium, where the quasi-statically driven model has a depinning transition at zero temperature.\cite{fisher98} In this case, it has been unambiguously proven through a perturbative functional renormalization group (FRG) treatment at two-loop order\cite{FRGledoussal-chauve} that the equilibrium and out-of-equilibrium critical behaviors are controlled by {\it distinct} fixed points. The RFIM on the other hand requires a nonperturbative FRG (NP-FRG) approach,\cite{footnote_perturbativeRG} which was first developed for studying the long-distance equilibrium behavior of the model and solve the puzzle of the dimensional-reduction breakdown and the associated supersymmetry breaking.\cite{tarjus04,tissier06,tissier11} 

In a previous attempt to generalize the NP-FRG formalism to the out-of-equilibrium (hysteresis) critical behavior,\cite{balog-tissier-tarjus} we predicted that the same fixed point controls equilibrium and hysteresis criticality. We show here that although the prediction is correct (and nontrivial) for dimensions $d$ above a critical value $d_{DR}\approx 5.1$ when the avalanche-generated cusps in the renormalized disorder cumulants are subdominant at long distance, it is wrong below $d_{DR}$. {\it For $d<d_{DR}\approx 5.1$ the equilibrium and out-of-equilibrium critical points of the RFIM are controlled by distinct fixed points and are thus in different universality classes}. We explain why the short-cut tried in Ref. [\onlinecite{balog-tissier-tarjus}] in the form of a static superfield formalism fails for describing the hysteresis behavior and must be replaced by a dynamical field theory that keeps track of the direction in which the applied field, hence the magnetization, changes. We generalize the NP-FRG approach to encompass this dynamical field theory, which allows us to make predictions for the out-of-equilibrium critical behavior. We propose that a clear-cut test of the differing criticalities is to investigate finite-size scaling functions that probe the breaking (for hysteresis) or not (for equilibrium) of the $Z_2$ inversion symmetry.

\section{Summary of results}
\label{summary}

We show that the equilibrium and hysteresis critical behaviors of the RFIM fall or not in the same universality class depending on the spatial dimension $d$. For high dimension, but still below the upper critical one,\cite{footnote_uppercritical} $d>d_{DR}\approx 5.1$, the equilibrium and out-of-equilibrium critical points are controlled by the same fixed point, as indicated in our previous work.\cite{balog-tissier-tarjus} This is so despite the fact that the system near criticality on either the ascending or the descending branch of the hysteresis curve has no $Z_2$ inversion symmetry in contrast with the system in equilibrium: the symmetry is then asymptotically restored at the end of the flow, {\it i.e.}, at the fixed point. The situation is however different in lower dimensions, for $d<d_{DR}$, which includes the physical dimension $d=3$. The equilibrium and out-of-equilibrium critical points are controlled by {\it different} fixed points and thus belong to {\it different} universality classes. In particular, the hysteresis fixed point no longer shows the $Z_2$ symmetry. 

The reason why the pattern is different above and below $d_{DR}\approx 5.1$ is that the cumulants of the renormalized random field develop a strong nonanalytic dependence on their (field) arguments (a ``cusp'') at the fixed point below $d_{DR}$. This cusp is generated by the avalanches that characterize the evolution of the system either in or out of equilibrium at zero temperature. (It also leads to the breaking of the $d \to (d-2)$ dimensional-reduction property and the spontaneous breaking of the underlying supersymmetry of the model.) When $d>d_{DR}$, avalanches are still present but lead to subdominant effects at large distance, and the cusp becomes irrelevant at the fixed point. The exact RG equations for the RFIM in and out of equilibrium are identical when there is no cusp but are different when a cusp is present (a mechanism that was overlooked in the static superfield approach used in our previous work:\cite{balog-tissier-tarjus} see the next section). A similar phenomenon takes place in the case of an elastic manifold, such as an interface, in a disordered environment: the presence of a cusp, which always occurs below the upper critical dimension in this case, leads to different RG equations for the equilibrium case and the out-of-equilibrium depinning problem at a high enough order of the perturbative, but functional,  treatment.\cite{FRGledoussal-chauve}

The hysteresis critical points are thus in a different universality class than the equilibrium one in $d\lesssim 5.1$. However, it was found in computer simulations of the RFIM in $d=3$ that the differences in the critical exponents and in the integrated avalanche distribution functions are small. This is also what our present theoretical approach shows: the exponents that we can determine with better accuracy, which are the anomalous dimensions characterizing the fixed points, are extremely close between hysteresis and equilibrium, despite the qualitative difference of the fixed points, showing or not the $Z_2$ symmetry. 

\begin{figure}[ht]
\begin{center}
\includegraphics[width=\linewidth]{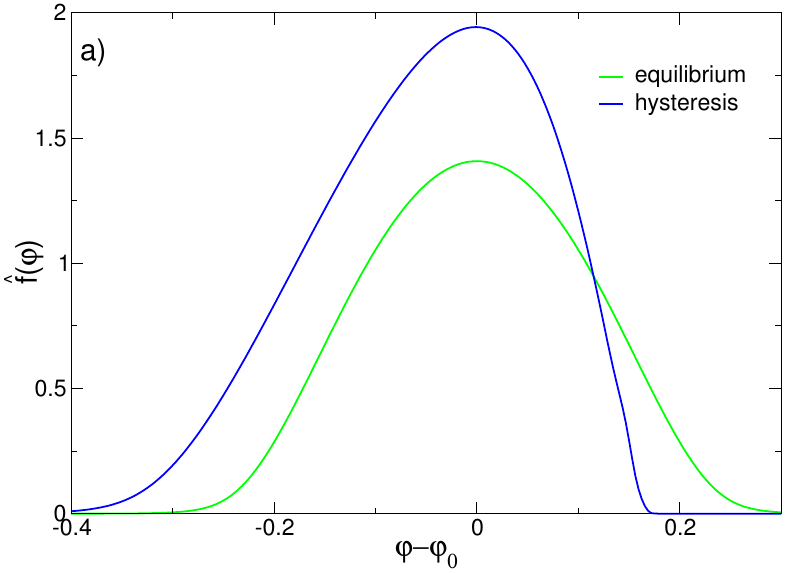}
\includegraphics[width=0.97\linewidth]{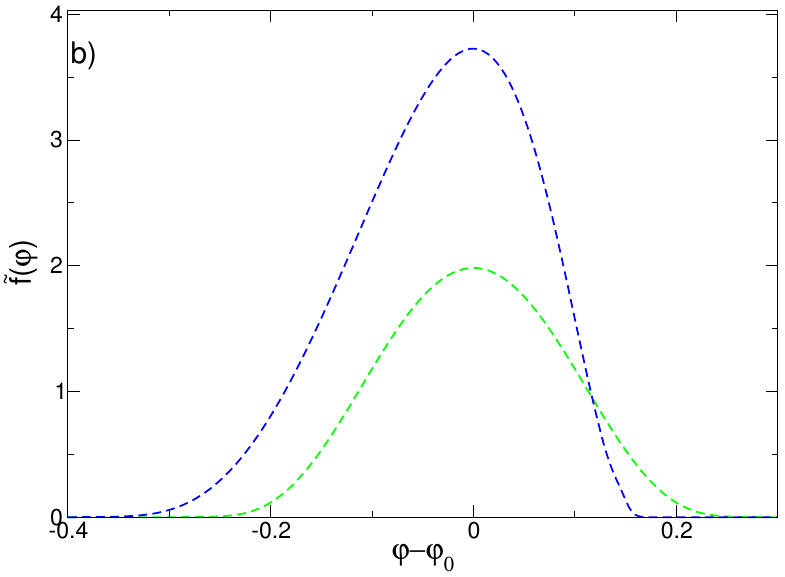}
\caption{Finite-size scaling functions $\hat f(\varphi)$ for the connected susceptibility (a) and $\tilde f(\varphi)$ for the disconnected susceptibility (b) for equilibrium and hysteresis criticality, as predicted by the present NP-FRG theory from fixed-point functions in $d=3.6$. (We consider here the ascending branch of the hysteresis loop.) For the hysteresis the maxima are located at a nonzero offset field $\varphi_0$, as a result of the lack of $Z_2$ symmetry, whereas for equilibrium $\varphi_0=0$. See section \ref{results} for more details.}
\label{Fig_FSS}
\end{center}
\end{figure}

We therefore propose that the only way to provide a clear-cut numerical confirmation of the differing critical behaviors in $d=3$ is to study the presence or not of $Z_2$ symmetry in scaling functions. The $Z_2$ symmetry cannot be seen in the integrated distribution of avalanche sizes which is commonly investigated\cite{perkovic,perez03,perez04,liu09} but is probed when approaching the critical point on, say, the ascending branch of the hysteresis curve from either higher or lower magnetic fields (or magnetizations).  One way to do this in a computer simulation is to work with a lattice of linear size $L$ at the critical value of the disorder empirically determined for this size, $\Delta_{c,L}$. After averaging over many samples, one can determine the effective location of the critical point $(H_{c,L}, m_{c,L})$ for the magnetic field and the mean magnetization per spin, {\it e.g.}, by locating where the magnetic (``connected'') susceptibility $\partial m/\partial H$ attains its maximum as a function of $H$. The quantities of interest are for instance this connected susceptibility $\widehat \chi_{c,L}(H)$ and the ``disconnected'' susceptibility $\widetilde \chi_{c,L}(H)=L^d[\overline{m_L(H)^2}-\overline{m_L(H)}^2]$, where $m_L(H)$ is the magnetization reached in a given sample by applying the $T=0$ quasi-static driving protocol when the magnetic field is increased to the value $H$. The scaling functions obtained by finite-size scaling collapses,
\begin{equation}
\begin{aligned}
\label{eq_finitesize_scaling}
&\hat f=  L^{-(2-\eta)} \widehat \chi_{c,L}([H-H_{c,L}]L^{\frac 12(d-2\eta+\bar\eta)}) \\&
\tilde f=  L^{-(4-\bar\eta)} \widetilde \chi_{c,L}([H-H_{c,L}]L^{\frac 12(d-2\eta+\bar\eta)})
\end{aligned}
\end{equation}
can also be plotted as a function of the reduced magnetization, $\varphi=L^{(d-4+\bar\eta)/2)}[\overline{m_L(H)}-m_{c,L}]$. These finite-size scaling functions $\hat f(\varphi)$ and $\tilde f(\varphi)$ can be related to fixed-point functions obtained by the NP-FRG, as will be detailed in the rest of the paper. We illustrate the behavior of the predicted functions for equilibrium and hysteresis in Fig. \ref{Fig_FSS} for $d=4$. (As explained below,  the lowest dimension we could access for hysteresis  is $d\approx 3.5$ due to the build-up of numerical difficulties in solving the needed coupled set of partial differential equations as $d$ decreases.)

\begin{figure}[ht]
\begin{center}
\includegraphics[width=\linewidth]{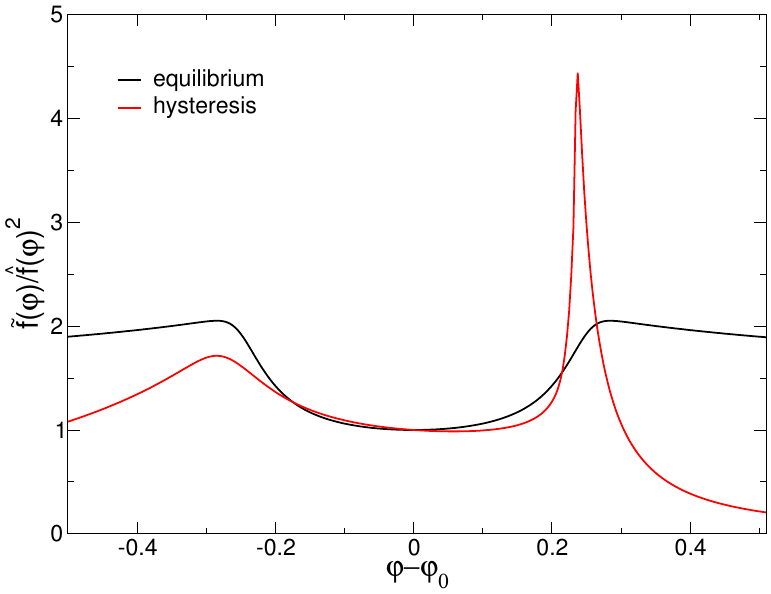}
\caption{Ratio of scaling functions $\tilde f(\varphi)/\hat f(\varphi)^2$ for equilibrium and hysteresis criticality, as predicted by the present NP-FRG theory from the fixed-point dimensionless second cumulant of the renormalized random field in $d=3.6$. See section \ref{results} for more details.}
\label{Fig_FSSratio}
\end{center}
\end{figure}

As can be seen, the finite-size scaling functions are predicted to be different for equilibrium and out-of-equilibrium criticality. In contrast with the equilibrium ones,  those for hysteresis show a detectable {\it asymmetry} around the maximum. This asymmetry is much more striking when plotting $\tilde f(\varphi)/\hat f(\varphi)^2$, which as discussed in sections \ref{flow_eqs} and \ref{results},  corresponds to the dimensionless second cumulant of the renormalized random field: see Fig. \ref{Fig_FSSratio}. The latter is directly obtained from the NP-FRG but might be hard to extract with a good precision from the finite-size scaling analysis of simulation data.
\\

\section{Deficiency of the static superfield approach: an erratum to our previous work}
\label{deficiency}

In an earlier work\cite{balog-tissier-tarjus} we aimed at replacing the \textit{a priori} complex problem of following a history-dependent evolution among configurations, which results from the dynamics of the slowly driven system, by one which is in effect ``static'' and history-independent. The limit of interest is indeed the quasi-static one in which the driving rate is vanishingly slow so that the system reaches a stationary state before being evolved again.\cite{sethna93} Due to the properties of the system and of the $T=0$ dynamics,\cite{sethna93} the configurations visited along the  process correspond to \textit{extremal} states:\cite{sethna93,balog-tissier-tarjus,guagnelli93,mlr10} for a given value of the applied external magnetic field, they correspond to the stationary states (``metastable states'') that have, at each point in space, the largest or smallest (depending on the branch which is followed) local magnetization. In principle these extremal states can therefore be selected with no reference to dynamics and history. Furthermore, when the distribution of the disorder is continuous, which we consider, the ground state as well as the extremal states are unique for a given realization of the disorder, with only exceptional degeneracies. 

We then tried to generalize the superfield formalism that we had previously successfully used to study the equilibrium RFIM at $T=0$ via its ground-state properties\cite{tissier11} to investigate the extremal states of the system. The key point was to be able to devise a proper selection of either the ground state or the extremal states among the multitude of metastable states of the system. 

The problem however that we overlooked in Ref.~[\onlinecite{balog-tissier-tarjus}] originates in the nonanalytic functional dependence (the linear ``cusp'') appearing in the cumulants of the renormalized random field as a result of the presence of large scale avalanches at zero temperature. As was first unveiled in the case of elastic manifolds in a disordered environment,\cite{FRGledoussal-chauve} the exact, functional, RG flow equations involve terms that are ambiguous when the equations are naively considered at $T=0$, because they require derivatives of the disorder cumulants in the direction of the cusp that are {\it a priori} singular. For the equilibrium case, this can be handled by considering the limiting process of taking a small temperature to zero, where in the superfield formalism this temperature is an auxiliary one that is used to select the ground state among all possible metastable states.\cite{tissier11} The presence of an infinitesimal $T$ indeed rounds the cusp, and any singular behavior associated with it, in a thermal boundary layer.\cite{chauve_creep,balents-ledoussal,ledoussal10,tissier06,tissier11,balog_activated} The limit $T=0$ inherits of the symmetry properties found in a rounded cusp at nonzero $T$ in the sense that both directions of the cusp are equivalent.

The situation is however different for the extremal states that are selected at $T=0$ via the introduction of large auxiliary sources coupled to the magnetization.\cite{balog-tissier-tarjus} These sources indeed {\it do not} round the cusp, and the exact FRG equations therefore always contain terms that are singular when a cusp is present. In Ref. [\onlinecite{balog-tissier-tarjus}] we therefore wrongly concluded that the flow equations for the equilibrium and out-of equilibrium RFIM are always identical. This property is only true for the range of dimensions $d>d_{DR}\approx 5.1$ where the cuspy behavior due to avalanches is subdominant at large distance.\cite{tarjus13} For $d<d_{DR}$ one cannot conclude from the superfield formalism. (In technical terms, in the hysteresis case, the ``corrections to Grassmannian ultralocality'' diverge in the presence of a cusp, a point that we missed in Ref. [\onlinecite{balog-tissier-tarjus}].)

There does not seem to be any shortcut via a static superfield formalism for studying critical hysteresis in the quasi-statically driven RFIM and, just like for the case of elastic manifolds in a disordered environment,\cite{FRGledoussal-chauve} one has to resort to a dynamical field theoretical treatment. We show below that this is indeed possible.

\section{Dynamical field theory for the RFIM in and out of equilibrium}
\label{dynamical}

As we are interested in the long-time collective behavior of the RFIM, a coarse-grained field theory provides an appropriate starting point. We therefore consider the following ``bare action'' (microscopic Hamiltonian) for a scalar field $\varphi$ in a $d$-dimensional space:
\begin{equation}
\begin{aligned}
\label{eq_ham_dis_RFIM}
&S_{dis}[\varphi;h]=  S_B[\varphi]-\int_{x} h_x \varphi_x \,, \\&
S_B[\varphi]= \int_{x}\bigg\{\frac{1}{2}(\partial_x \varphi_x)^2+ \frac{r}{2} \varphi_x^2 + \frac{u}{4!} \varphi_x^4 \bigg\},
\end{aligned}
\end{equation}
where $ \int_{x} \equiv \int d^d x$ and $h_x$ is a random ``source'' (a random magnetic field); this quenched random field $h$ is taken with a Gaussian distribution characterized by a zero mean and  a variance $\overline{h_xh_y}= \Delta_B \delta^{(d)}( x-y)$ (corresponding to short-ranged correlations). An ultraviolet (UV) cutoff $\Lambda$ on the momenta, associated with the inverse of a microscopic length scale such as a lattice spacing, is also implicitly taken into account.

At a coarse-grained level, the dynamics of the system both near to and far from equilibrium can be described by a Langevin equation,
\begin{equation}
\label{eq_stochastic_dynamics_RFIM}
\partial_t\varphi_{xt}=-\frac{\delta S_B[\varphi]}{\delta \varphi_{xt}}+h(x)+J_{xt} + \eta_{xt},
\end{equation}
where $\eta_{xt}$ is a Gaussian random thermal noise with zero mean and variance $\langle\eta_{xt}\eta_{x't'}\rangle=2T \delta^{(d)}(x-x')\delta(t-t')$, and $J_{xt}$ is the applied source (or magnetic field). The relaxation dynamics to equilibrium corresponds to taking $T>0$ and $J$ independent of time. Actually, due to the statistical $Z_2$ inversion symmetry, the critical point then takes place for $J=0$. On the other hand, the quasi-statically driven situation leading to hysteresis corresponds to $T=0$ and $J_{xt}=J+\Omega t$ with $\Omega\to 0^+$ on the ascending branch of the hysteresis loop and $\Omega\to 0^-$ on the descending one. The critical point on the ascending branch takes place at $J_c>0$ and that on the descending branch at $-J_c$.

The generating functional of the multi-point and multi-time correlation and response functions can be built as usual by following the Martin-Siggia-Rose-Janssen-de Dominicis formalism.\cite{MSR,janssen-dedom} After introducing an auxiliary ``response'' field $\hat \varphi_{xt}$ and taking into account the fact that the solution of Eq. (\ref{eq_stochastic_dynamics_RFIM}) is unique,\cite{zinnjustin89} one obtains the generalized ``partition function''
\begin{equation}
\begin{aligned}
\label{eq_part_dis2}
\mathcal Z_{h,\eta}[\hat{J},J]=&\int \mathcal{D}\varphi\mathcal{D}\hat{\varphi} \exp\big\{-\int_{xt}\hat{\varphi}_{xt}\Big[ \partial_t\varphi_{xt}+ \frac{\delta S_B[\varphi]}{\delta\varphi_{xt}}  \\& 
-\eta_{xt} -h_x\Big]+\int_{xt}(\hat{J}_{xt}\varphi_{xt}+J_{xt}\hat{\varphi}_{xt})\Big \}\,
\end{aligned}
\end{equation}
where we have used the It\={o} prescription (which amounts to setting to 1 the Jacobian of the transformation between the thermal noise and the field).\cite{zinnjustin89} 

To handle the averages over the thermal noise and the random field we follow the same procedure as in Ref. [\onlinecite{balog_activated}]. Keeping in mind that the key point for taking relevant events such as avalanches and droplets into account is to describe the full functional dependence of the cumulants of the renormalized disorder,\cite{tarjus04,tissier06,tissier11} we introduce copies or replicas of the system:  the copies have the {\it same} disorder $h$ but are coupled to \textit{distinct} sources and are characterized by {\it independent} thermal noises. Contrary to the conventional practice,\cite{dedom78} we therefore combine  dynamics \textit{and} replicas.

After averaging over the thermal noises and the disorder, one obtains
\begin{equation}
\begin{aligned}
\label{eq_part_aver}
&Z[{\hat{J}_a,J_a}]= \\& \int \prod_a \mathcal{D}\varphi_a\mathcal{D}\hat{\varphi}_a e^{-S_{dyn}[\{\hat{\varphi}_a,\varphi_a\}]+\sum_a\int_{xt}(\hat{J}_{a,xt}\varphi_{a,xt}+J_{a,xt}\hat{\varphi}_{a,xt})}
\end{aligned}
\end{equation}
where the (bare) dynamical action is
\begin{equation}
\begin{aligned}
\label{eq_bare_action}
S_{dyn}[\{\hat{\varphi}_a,&\varphi_a\}]= 
\sum_a \int_{xt}\hat{\varphi}_{a,xt}\Big\{\partial_t\varphi_{a,xt}-T\hat{\varphi}_{a,xt}
\\&  +\frac{\delta S_B[\varphi_a]}{\delta{\varphi}_{a,xt}}\Big\} -\frac{\Delta_B}{2}\sum_{ab}\int_{xtt'}  \hat{\varphi}_{a,xt}\hat{\varphi}_{b,xt'}\,.
\end{aligned}
\end{equation}
Note the difference between the effect of the thermal noises that lead to a one-replica term which is local both in space and in time and the effect of the quenched random field that generates a two-replica term which is local in space but not in time.\cite{balog_activated}

Causality comes with It\={o}'s prescription (which will be further discussed below) and should apply for both the equilibrium and the hysteretic dynamics. In addition, the relaxation toward equilibrium satisfies an invariance under time translation and a time-reversal symmetry.\cite{zinnjustin89} The latter in turn implies the fluctuation-dissipation theorem relating correlation and response functions.\cite{janssen92,zinnjustin89,andreanov06} The time-reversal symmetry corresponds to an invariance of the theory under the simultaneous transformations\cite{andreanov06} 
\begin{equation}
\begin{aligned}
&t \rightarrow -t\,,\\&
\varphi_a \rightarrow \varphi_a\, , \\&
\hat\varphi_a \rightarrow \hat\varphi_a-(1/T)\partial_t\varphi_a\, .
\end{aligned}
\end{equation}
It applies only for $T>0$ and is therefore {\it a priori} violated in the out-of-equilibrium, driven case.

\section{Nonperturbative functional renormalization group}
\label{NPFRG}

\subsection{The effective average action formalism}

To proceed further, we apply the nonperturbative functional renormalization group (NP-FRG) to describe the long-distance and long-time physics of this dynamical theory near criticality.  As in Ref. [\onlinecite{balog_activated}] we do so by generalizing the formalism developed for the (static) equilibrium properties of the RFIM\cite{tarjus04,tissier06,tissier11} and combining it with the approach put forward for the critical dynamics of the Ising model in the absence of quenched disorder.\cite{canet} The exact renormalization group procedure amounts to progressively including the contribution of the fluctuations of the dynamical field on longer length and time scales, or alternatively, with shorter momenta and frequencies.\cite{wilson74,wetterich93,berges02} 

Technically, the procedure can be implemented through the addition to the bare action of an ``infrared (IR) regulator'' $\Delta S_k$ depending on a running IR scale $k$. Its role is to suppress the integration over slow modes associated with momenta $\vert q \vert \lesssim k$ in the functional integral.\cite{wetterich93,berges02,tarjus04,tissier11} We therefore add to the replicated dynamical action in Eq. (\ref{eq_bare_action})
\begin{equation}
\begin{aligned}
\label{bare_action}
&\Delta S_{k}[\{\Phi_a\}]= \\&\frac 12 \int_{xt}\int_{x't'}{\rm{tr}}\Big[\sum_a \Phi_{a,xt}\widehat{\mathbf R}_k(x-x',t-t') \Phi_{a,x't'}^\top \\&
+\frac 12 \sum_{ab} \Phi_{a,xt}\widetilde{\mathbf R}_k(x-x',t-t') \Phi_{b,x't'}^\top\Big] \,,
\end{aligned}
\end{equation}
where $\Phi_a\equiv (\varphi_a,\hat\varphi_a)$,  $\Phi_a^\top$ is its transpose, and the trace is over the 2 components of $\Phi_a$; $\widehat{\mathbf R}_k$ and $\widetilde{\mathbf R}_k$ are symmetric $2\times 2$ matrices of mass-like IR cutoff functions that enforce the decoupling between fast (high-momentum) and slow (low-momentum) modes in the partition function. Following Ref. [\onlinecite{canet}] it proves sufficient to control the contribution of the fluctuations through their momentum dependence and we take
\begin{equation}
\begin{aligned}
\label{hatR}
\widehat R_{k,11}=\widehat R_{k,22}=0 \;,  \widehat R_{k,12}=\widehat R_{k,21}=\widehat R_k(\vert x-x'\vert)\,,
\end{aligned}
\end{equation}
and 
\begin{equation}
\begin{aligned}
\label{tildeR}
\widetilde R_{k,11}=\widetilde R_{k,12}=\widetilde R_{k,21}=0 \;, \widetilde R_{k,22}=\widetilde R_k(\vert x-x'\vert)\,,
\end{aligned}
\end{equation}
where  the Fourier transforms, $\widehat R_k(q^2)$ and $\widetilde R_k(q^2)$, are chosen  such that the integration over modes with momentum $\vert q \vert \lesssim k$ is suppressed.\cite{berges02,tarjus04,tissier11} To avoid an explicit breaking of the underlying supersymmetry (invariance under rotations in superspace) of the theory,\cite{parisi79} we take\cite{tissier11}
\begin{equation}
\label{eq_ward_surot}
\widetilde R_k(q^2)\propto \frac{\partial \widehat R_k(q^2)}{\partial q^2}\, .
\end{equation}
Note that the above choice of IR regulator satisfies both the time-reversal and the $Z_2$ symmetry, $\{\Phi_a\} \to \{-\Phi_a\}$.

The partition function $Z[\{\mathcal J_a\}]$, where $\mathcal J_a$ denotes $(\hat J_a,J_a)$ is then replaced by a $k$-dependent quantity,\begin{equation}
\begin{aligned}
\label{eq_part_aver_k}
&Z_k[\{\mathcal J_a\}]= \\& \int \prod_a \mathcal{D}\Phi_a e^{-S_{dyn}[\{\Phi_a\}]+\sum_a\int_{xt} \mathcal{J}_{a,xt} \Phi_{a,xt}^\top-\Delta S_{k}[\{\Phi_a\}]}\,.
\end{aligned}
\end{equation}

The  central quantity of our NP-FRG is the ``effective average action" $\Gamma_k$,\cite{wetterich93} which is the generating functional of the 1-particle irreducible (1PI) correlation functions at the scale $k$. It is defined (modulo the subtraction of a regulator contribution) from $\ln Z_k[\{\mathcal J_a\}]$ via a Legendre transform:
\begin{equation}
\label{eq_legendre_transform_k}
\Gamma_k[\{\Phi_a\}]+\ln{Z_k[\{\mathcal J_a\}]}= \sum_a \int_{xt} {\rm{tr}} \mathcal{J}_{a,xt}\Phi_{a,xt}^\top-\Delta S_k[\{\Phi_a\}],
\end{equation}
where $\Phi_a\equiv (\phi_a,\hat\phi_a)$ now denotes the ``classical'' (or average) fields with 
\begin{equation}
\begin{aligned}
\label{eq_legendre}
&\phi_{a,xt}=\frac{\delta \ln Z_k}{\delta \hat J_{a,xt}}=\langle \varphi_{a,xt}\rangle
\\&
\hat{\phi}_{a,xt}=\frac{\delta \ln Z_k}{\delta J_{a,xt}}=\langle \hat\varphi_{a,xt}\rangle\,.
\end{aligned}
\end{equation}
The trace in Eq. (\ref{eq_legendre_transform_k}) is over the 2 components of $\Phi_a$ and $\mathcal J_a$. 

The effective average action $\Gamma_k$ satisfies an {\it exact} RG equation (ERGE) describing its evolution with the IR cutoff $k$,\cite{wetterich93}
\begin{equation}
\label{eq_wetterich}
\partial_k\Gamma_k[\{\Phi_a\}]=\frac 12 \rm{Tr} \big\{ (\partial_k\mathbf R_k) (\bm\Gamma_k^{(2)}[\{\Phi_a\}] + \mathbf R_k)^{-1} \big\},
\end{equation}
where the trace is over space-time coordinates, copy indices and
components, and $\bm\Gamma_k^{(2)}$ is the matrix formed by the second functional derivatives of $\Gamma_k$. (In what follows, superscripts within a parenthesis are used to indicate derivatives with respect to the appropriate arguments.)

\subsection{NP-FRG for the cumulants of the renormalized random field}

We are interested in the cumulants of the renormalized disorder. The (generalized) cumulants at the scale $k$ can be generated by expanding the effective average action in increasing number of unrestricted sums over copies,\cite{tarjus04,tissier06,tissier11}
\begin{equation}
\label{eq_expansion_gamma_k}
\Gamma_k[\{\Phi_a\}]=\sum_{p=1}^\infty \frac {(-1)^{p-1}}{p!}\sum_{a_1\cdots a_p} \mathsf \Gamma_{kp}[\Phi_{a_1},\cdots,\Phi_{a_p}]
\end{equation}
where, as a result of causality and It\=o's prescription,\cite{footnote_ito} $\mathsf \Gamma_{kp}$ can be put in the following form:
\begin{equation}
\begin{aligned}
\label{eq_gamma_kp}
\mathsf \Gamma_{kp}=\int_{x_1t_1}\cdots \int_{x_p t_p}\hat\phi_{a_1,x_1t_1}\cdots \hat\phi_{a_p,x_pt_p} \gamma_{k p;x_1t_1,\cdots,x_pt_p}
\end{aligned}
\end{equation}
with $\gamma_{k p;x_1t_1,\cdots,x_pt_p}$ a functional of the fields  $\Phi_{a_1,t_1},\cdots, \Phi_{a_p,t_p}$ and of their time derivatives, $\partial_{t_1}^q \Phi_{a_1,t_1},\cdots, \partial_{t_p}^q \Phi_{a_p,t_p} $, $q\geq 1$. Note that the fields and their time derivatives are taken at fixed time values $t_1,\cdots,t_p$ whereas space points are not specified. When the fields are chosen uniform in time with $\phi_{a,xt}=\phi_{a,x}$ and $\hat\phi_{a,xt}=0$, the $\gamma_{k p}$'s reduce to the cumulants of the renormalized random field already studied at equilibrium in Refs. [\onlinecite{tarjus04,tissier06,tissier11}]. For generic ({\it i.e.}, nonuniform in time) fields, the additional contributions represent kinetic terms; {\it e.g.},\cite{balog_activated,balents-ledoussal_dyn} for the first cumulant,
\begin{equation}
\begin{aligned}
\label{eq_kinetic terms}
&\gamma_{k 1;x_1t_1}[\Phi_{a_1,t_1};\partial_{t_1} \Phi_{a_1,t_1};\partial_{t_1}^2 \Phi_{a_1,t_1};\cdots]=  \\& 
 \sum_{m,n\geq 0}\, \sum_{q,r \geq 1}\gamma_{k 1;\{m,q\},\{n,r\}}[\Phi_{a_1,t_1}](\partial_{t_1}^m \phi_{a_1,t_1})^q(\partial_{t_1}^n\hat\phi_{a_1,t_1})^r
 \end{aligned}
\end{equation}
where we have dropped the explicit dependence on the subscript $x_1$ on all the functionals appearing in the right-hand side. The functionals can be further expanded in powers of $\hat\phi_{a_1,t_1}$ around zero.\cite{balents-ledoussal_dyn} 

One can then distinguish between the {\it nonlocal} time dependence associated with the indices $t_1,\cdots,t_p$ corresponding to the times that are integrated over without restriction in Eq. (\ref{eq_gamma_kp}) and the {\it quasi-local} time dependence that amounts to considering the fields and their time derivatives at the same value of time and can be handled through a derivative expansion as in Eq.~(\ref{eq_kinetic terms}).

 After inserting the expansion in increasing number of sums over copies, Eq. (\ref{eq_expansion_gamma_k}), in the ERGE for the effective average action, Eq. (\ref{eq_wetterich}),  and proceeding to algebraic manipulations, one can derive an infinite hierarchy of coupled ERGE's for the generalized cumulants $\mathsf \Gamma_{kp}$. For instance, with the choice of IR regulator in Eqs.~(\ref{hatR},\ref{tildeR}) the two first flow equations explicitly read
\begin{equation}
\label{eq_flow_Gamma1}
\begin{aligned}
&\partial_k \mathsf \Gamma_{k 1}\left[ \Phi_1\right ]=
\dfrac{1}{2} \int_{x_1x_2}  \int_{t_1}{\rm{tr}} \Big [\partial_k \widehat{\mathbf R}_k(\vert x_1-x_2\vert) \times \\& \big (\widehat{\mathbf P}_{k;(x_1 t_1)(x_2 t_1)}^{[0]}[ \Phi_1] +\widetilde{\mathbf P}_{k;(x_1t_1)(x_2 t_1)}^{[0]}[ \Phi_1, \Phi_1] \big )\, + \\&  \partial_k \widetilde {\mathbf R}_k(\vert x_1-x_2\vert) 
\widehat{\mathbf P}_{k;(x_1 t_1)(x_2 t_1)}^{[0]}[\Phi_1] \Big ]
\end{aligned}
\end{equation}
\begin{equation}
\label{eq_flow_Gamma2}
\begin{aligned}
&\partial_k \mathsf \Gamma_{k2}\left[ \Phi_1 , \Phi_2\right ]= \dfrac{1}{2} \widetilde{\partial}_k \bigg \{ \int_{x_3x_4} \int_{t_3t_4} {\rm{tr}}
 \Big [ \widehat{\mathbf P}_{k;(x_3 t_3)(x_4 t_4)}^{[0]}\left[ \Phi_1 \right ] 
 \\& ( \bm{\mathsf \Gamma} _{k2;(x_4 t_4)(x_3 t_3),.}^{(20)}\left[ \Phi_1,  \Phi_2 \right ] - \bm{\mathsf \Gamma} _{k3;(x_4 t_4)(x_3 t_3),.}^{(110)}\left[ \Phi_1,  \Phi_1,  \Phi_2 \right ])  
\\& + \widetilde{\mathbf P}_{k;(x_3t_3) (x_4t_4)}^{[0]}\left[ \Phi_1, \Phi_1 \right ] 
\bm{\mathsf \Gamma}_{k2;(x_4 t_4)(x_3 t_3),.}^{(20)}\left[ \Phi_1, \Phi_2 \right ] 
 + \\&
 \dfrac{1}{2} \widetilde{ \mathbf P}_{k;(x_3t_3)(x_4 t_4)}^{[0]}\left [ \Phi_1, \Phi_2 \right ]\Big (\bm{\mathsf \Gamma}_{k2;(x_4 t_4)(x_3 t_3)}^{(11)}\left[ \Phi_1, \Phi_2 \right ] \\&
 - \widetilde{\mathbf R}_{k}(\vert x_3-x_4\vert) \Big ) + perm (12) \Big ]\bigg \} \,,
\end{aligned}
\end{equation}
where we recall that the superscripts within parentheses on the $\mathsf \Gamma_{k p}$'s indicate functional derivatives. In the second equation we have introduced the short-hand notation $\widetilde{\partial}_k$ to indicate a derivative acting only on the cutoff functions (\textit{i.e.},  $\widetilde{\partial}_k \equiv \partial_k \widehat{R}_k\, \delta/\delta \widehat{R}_k + \partial_k \widetilde{R}_k \, \delta/\delta \widetilde{R}_k$) and $perm (12)$ denotes the expression obtained by permuting $ \Phi_1$ and $ \Phi_2$. Finally, $\widehat{\mathbf P}_{k}$ and $\widetilde{\mathbf P}_{k}$ are the components of the propagator, with $\mathbf P_{k,ab}=\delta_{ab} \widehat{\mathbf P}_{k,a} + \widetilde{\mathbf P}_{k,ab}$ and in an operator sense $\mathbf P_{k}=(\bm{\Gamma}_k^{(2)}+ \mathbf R_k)^{-1}$,  and the superscript $[0]$ indicates the lowest order of the expansion in increasing number of sums over copies. These propagators can be expressed in terms of second derivatives of the cumulants $\mathsf \Gamma_{k1}$ and $\mathsf \Gamma_{k2}$ as follows:
\begin{equation}
\label{eq_hatP_zero}
\mathbf{\widehat {P}}_{k;x_1t_1,x_2t_2}^{[0]}[\Phi ]=\left(\bm{\mathsf\Gamma}_{k1}^{(2)}[ \Phi ]+\mathbf{\widehat R}_k\right) ^{-1}\Big \vert_{x_1t_1,x_2t_2}
\end{equation}
and
\begin{equation}
\begin{aligned}
\label{eq_tildeP_zero}
&\mathbf{\widetilde {P}}_{k;x_1t_1,x_2t_2}^{[0]}[\Phi_1, \Phi_2 ]= \int_{x_3t_3}\int_{x_4t_4}\mathbf{\widehat {P}}_{k;x_1t_1,x_3t_3}^{[0]}[ \Phi_1 ] \times \\&
\Big (\bm{\mathsf\Gamma}_{k2;x_3t_3,x_4t_4}^{(11)}[\Phi_1, \Phi_2 ]
- \mathbf{\widetilde R}_k(\vert x_3-x_4\vert)\Big ) \mathbf{\widehat {P}}_{k;x_4t_4,x_2t_2}^{[0]}[ \Phi_2 ]\,.
\end{aligned}
\end{equation}

\section{Nonperturbative approximation scheme}
\label{approx}

The hierarchy of ERGE's cannot be solved exactly. We therefore consider the nonperturbative approximation scheme for the effective average action that was first introduced in the context of the RFIM at and near equilibrium.\cite{tarjus04,tissier11,balog_activated} It combines a truncation in the \textit{spatial derivative expansion}, \textit{i.e.} an expansion in the number of spatial derivatives for approximating the long-distance behavior of the 1PI correlation functions, a truncation in the  \textit{time derivative expansion and expansion in the auxiliary fields $\hat \phi_a$}, which amounts to truncating the number of kinetic coefficients for describing the long-time behavior,\cite{balents-ledoussal_dyn,gorokhov_creep} and a truncation in the \textit{expansion in cumulants of the renormalized disorder}. The scheme also ensures that the symmetries and supersymmetries of the theory are not explicitly violated, which turns out to be an important issue for a proper description of  dimensional reduction and its breakdown.\cite{tissier11}  The minimal truncation that already contains the long-distance and long-time physics of the RFIM both in and out of equilibrium (and is able to predict a difference of behavior between the two situations when there is one) can then be formulated as
\begin{equation}
\begin{aligned}
\label{eq_gammak1_ansatz}
\mathsf \Gamma_{k 1}[\Phi]=&\int_{xt}\hat\phi_{xt}\Big [J_k(\phi_{xt})+ \frac{1}{2}\frac{\delta}{\phi_{xt}}\big[Z_k(\phi_{xt})(\partial_x\phi_{xt})^2 \big ] \\& + Y_k(\phi_{xt})\partial_t\phi_{xt}-T X_k(\phi_{xt})\hat{\phi}_{xt}\Big ]
\end{aligned}
\end{equation}
\begin{equation}
\begin{aligned}
\label{eq_gammak2_ansatz}
\mathsf{\Gamma}_{k2}[\Phi_1,\Phi_2]=\int_{x}\int_{t_1t_2}\hat{\phi}_{1,xt_1}\hat{\phi}_{2,xt_2}\Delta_k(\phi_{1,xt_1},\phi_{2,xt_2})
\end{aligned}
\end{equation} 
\begin{equation}
\begin{aligned}
\label{eq_gammak3_ansatz}
\mathsf{\Gamma}_{kp}=0,\; \forall p\geq 3\,.
\end{aligned}
\end{equation} 
In the above equations, $J_k(\phi)$ is a renormalized force or source which in equilibrium is the derivative of the effective average potential, $J_k(\phi)=U_k'(\phi)$. In the hysteresis case, the effective potential has no physical meaning, which is why we use $J_k$ in place of $U_k'$. $Z_k(\phi)$ plays the role of a renormalization function for the field. The two kinetic functions $Y_k(\phi)$ and $X_k(\phi)$ are equal when the time-reversal symmetry applies (for $T>0$) and $X_k$ is irrelevant when describing the driven case at $T=0$. Finally, $\Delta_k(\phi_{1},\phi_{2})$ is the second cumulant of the renormalized random field. These functions are defined from the following exact prescriptions:
\begin{equation}
\begin{aligned}
 \label{eq_prescriptions}
 & J_k(\phi)= \frac{\delta \mathsf \Gamma_{k 1}[\Phi]}{\delta \hat\phi_{xt}}\Big\vert_{unif}\\&
 Z_k(\phi)=\lim_{p\to 0}\frac{d}{dp^2}FT_{p}\frac{\delta^2\mathsf{\Gamma}_{k1}[\Phi]}{\delta\phi_{xt}\delta\hat{\phi}_{x't}}\Big\vert_{unif}\\&
 Y_k(\phi)=\lim_{\omega\to 0}\frac{d}{d(-i\omega)} FT_{\omega} \frac{\delta^2\mathsf{\Gamma}_{k1}[\Phi]}{\delta\phi_{xt}\delta\hat{\phi}_{xt'}}\Big\vert_{unif}\\&
 T X_k(\phi)=-\frac 12 \frac{\delta^2\mathsf{\Gamma}_{k1}[\Phi]}{\delta\hat{\phi}_{xt}^2}\Big\vert_{unif}\\&
 \Delta_k(\phi_{1},\phi_{2})\delta^{(d)}(x_1-x_2)=\frac{\delta^2\mathsf{\Gamma}_{k2}[\Phi_1,\Phi_2]}{\delta\hat{\phi}_{1,xt_1}\delta\hat{\phi}_{2,xt_2}}\Big\vert_{unif}.
\end{aligned}
\end{equation}
where $FT$ means a Fourier transform and the subscript $unif$ means that we take configurations of the fields that are spatially uniform with $\phi_{a,xt}=\phi_{a,t}$ and $\hat\phi_{a,xt}=0$; the configurations are also either strictly uniform, {\it i.e.} constant, in time (for equilibrium) or very slowly evolving in an appropriate quasi-static limit (for hystesis): see below for a more detailed characterization.

After inserting the ansatz provided by Eqs. (\ref{eq_gammak1_ansatz}-\ref{eq_gammak3_ansatz}) in the ERGE's, Eqs. (\ref{eq_flow_Gamma1},\ref{eq_flow_Gamma2}), and using the RG prescriptions in Eq. (\ref{eq_prescriptions}), one obtains in principle a closed set of coupled nonperturbative functional RG equations for the functions $J_k$, $Z_k$, $Y_k$, $X_k$, and $\Delta_k$. There are however a number of conceptual difficulties to overcome before arriving at the final form of the flow equations. One comes from the need to implement It\=o's prescription in a nonperturbative setting to guarantee causality. This point was solved in Ref. [\onlinecite{canet}]. The other, more serious, problem is associated with the appearance of a nonanalytic field dependence in the cumulants of the renormalized random field when working at $T=0$. Before delving more into the solution of these difficulties, we first derive the expressions for the propagators and the 1PI vertices that are needed in the flow equations.

\section{Propagators and vertices}
\label{propagators}

The ``hat propagator'' at zeroth order of the expansion in free replica sums, $\mathbf{\widehat {P}}_{k}^{[0]}[\Phi ]$, is a key building block of the formalism. In all the RG flow equations resulting from the above nonperturbative ansatz it appears in field configurations that are uniform in space with moreover $\hat\phi=0$. It can be obtained from the second derivative of the approximate first cumulant expression given in Eq. (\ref{eq_gammak1_ansatz}) via Eq. (\ref{eq_hatP_zero}). For spatially uniform fields $\phi_{xt}=\phi_t$ and $\hat\phi_{xt}=0$, one finds
\begin{equation}
\begin{aligned}
\label{gamma1_2}
&\bm{\mathsf\Gamma}_{k1;x_1t_1,x_2t_2}^{(2)}[\Phi]=  \\&
\Bigg( \begin{array}{cc}
0 &C(\phi_{t_2},\partial_{t_2}\phi_{t_2},\partial_{x_2}^2,\partial_{t_2})\\
\\
C(\phi_{t_1},\partial_{t_1}\phi_{t_1},\partial_{x_1}^2,\partial_{t_1})&- 2T X_k(\phi_{t_1}) \end{array} \Bigg) \\&
\qquad \times \delta^{(d)}(x_1-x_2) \delta(t_1-t_2) 
\end{aligned}
\end{equation}
where $C(\phi_{t_a},\partial_{t_a}\phi_{t_a},\partial_{x_a}^2,\partial_{t_a})=J'_k(\phi_{t_a})-Z_k(\phi_{t_a}) \partial_{x_a}^2 +Y_k(\phi_{t_a}) \partial_{t_a}+Y'_k(\phi_{t_a}) (\partial_{t_a}\phi_{t_a})$, with $a=1,2$, is an operator acting on $\delta^{(d)}(x_1-x_2) \delta(t_1-t_2)$. 

Since the equilibrium case has already been studied in detail in our previous papers,\cite{tissier11,tarjus-balog13,balog_activated} we focus now on the hysteresis case and will stress the differences with the dynamical treatment of the RFIM at and near equilibrium. As we have emphasized, relaxation to equilibrium corresponds to taking $T>0$ and the driving rate $\Omega=0$. The limit to $T=0$ can then be taken and requires a careful account of the nonuniform convergence associated with the presence of a thermal boundary layer.\cite{balents-ledoussal,ledoussal10,tissier06,tissier11,balog_activated} On the other hand, out-of-equilibrium hysteresis criticality corresponds to setting first $T=0$ and considering the quasi-static limit of $\Omega\to 0^+$ for the ascending branch of the hysteresis curve or $\Omega\to 0^-$ for the descending one. As discussed by Dahmen and Sethna,\cite{dahmen96} an infinitesimal driving rate translates into an infinitesimal velocity for the magnetization, so that we should consider configurations such that
\begin{equation}
\label{eq_driving_velocity}
\phi_t=\phi + vt,\; v\to 0^\pm
\end{equation}
where the plus sign is for the ascending branch and the minus for the descending one.

For the hysteresis case [when $\phi_t$ is given by Eq. (\ref{eq_driving_velocity})] and the equilibrium case [when $\phi_t$ is constant], one can Fourier transform over space and time coordinates  the second derivative of the first cumulant, from which one obtains the hat propagator as
\begin{equation}
\begin{aligned}
\label{hatpropagator}
&\mathbf{\widehat {P}}_{k}^{[0]}(p^2,\omega;\Phi)=  
\Bigg( \begin{array}{cc}
G_k(p^2,\omega;\phi) \;& \; G_k^+(p^2,\omega;\phi)\\
\\
G_k^-(p^2,\omega;\phi) \; &\; 0 \end{array} \Bigg) + {\rm O}(v)
\end{aligned}
\end{equation}
where  
\begin{equation}
G_k^+(p^2,\omega;\phi)=[J'_k(\phi)+Z_k(\phi)p^2+\widehat R_k(p^2) - i\omega Y_k(\phi)]^{-1}
\end{equation} 
is the response function at the scale $k$, $G_k^-(p^2,\omega;\phi)=G_k^+(p^2,-\omega;\phi)$, and
\begin{equation}
G_k(p^2,\omega;\phi)=2 T X_k(\phi)G_k^+(p^2,\omega;\phi)G_k^-(p^2,\omega;\phi)
\end{equation} 
is the correlation function at the scale $k$.

Transforming back to time leads to
\begin{equation}
\begin{aligned}
\label{response_function}
&G_k^+(p^2,t',t;\phi) \equiv \int_x e^{ip(x-x')}\langle\hat\varphi_{x't'}\varphi_{xt} \rangle_k \\&
=\frac 1{Y_k(\phi)} e^{-\frac{J'_k(\phi) + Z_k(\phi)p^2+\widehat R_k(p^2)}{Y_k(\phi)}(t-t')}\theta(t-t')\,,
\end{aligned}
\end{equation}
with $\theta$ the Heaviside step function, and  $G_k^-(p^2,t',t;\phi)=G_k^+(p^2,t,t';\phi)$. The correlation function on the other hand is equal to
\begin{equation}
\begin{aligned}
&G_k(p^2,t',t;\phi) \equiv \int_x e^{ip(x-x')}\langle \varphi_{x't'}\varphi_{xt} \rangle_k \\&
=T \frac {X_k(\phi) \,e^{-\frac{J'_k(\phi) + Z_k(\phi)p^2+\widehat R_k(p^2)}{Y_k(\phi)}\vert t-t' \vert}}{Y_k(\phi)[J'_k(\phi) + Z_k(\phi)p^2+\widehat R_k(p^2)]}
\end{aligned}
\end{equation}
and is of course equal to zero when $T=0$ (in particular for the hysteresis case).

It\=o's prescription is enforced in the nonperturbative RG by ensuring that the response function $G_k^+(t',t)$ is zero when the two times coincide. This is achieved by everywhere shifting the time for the auxiliary response field by an infinitesimal positive amount:  $\langle\hat\varphi_{x't'}\varphi_{xt} \rangle_k \to \langle\hat\varphi_{x't'+\epsilon}\varphi_{xt} \rangle_k$ with $\epsilon \to 0^+$.\cite{canet}

We now turn to the ``tilde propagator'' at zeroth order of the expansion in free replica sums, $\mathbf{\widetilde {P}}_{k}^{[0]}[\Phi ]$. Its expression requires the second derivative of the second cumulant [see Eq. (\ref{eq_tildeP_zero})], which from the ansatz in Eq. (\ref{eq_gammak2_ansatz}) is obtained as
\begin{equation}
\begin{aligned}
\label{gamma2_11}
&\bm{\mathsf\Gamma}_{k2;x_1t_1,x_2t_2}^{(11)}[\Phi_1, \Phi_2 ]= \delta^{(d)}(x_1-x_2) \times \\&
\Bigg( \begin{array}{cc}
\hat\phi_{1}\hat\phi_{2}\Delta_k^{(11)}(\phi_{1},\phi_{2}) \;&\; \hat\phi_{1}\Delta_k^{(10)}(\phi_{1},\phi_{2})\\
\\
\hat\phi_{2}\Delta_k^{(01)}(\phi_{1},\phi_{2}) \; &\; \Delta_k(\phi_{1},\phi_{2}) \end{array} \Bigg)
\end{aligned}
\end{equation}
where we have dropped the explicit dependence of the fields on space and time coordinates in the matrix, {\it i.e.}, $\phi_a \to \phi_{a,x_at_a}$ and same for $\hat\phi_a$. When evaluated for configurations such that $\hat\phi_a=0$, all terms in the matrix vanish except the lower right one. The tilde propagator has then only one nonzero component, the upper left one, which for spatially uniform fields is then simply given in Fourier space by 
\begin{equation}
\begin{aligned}
\label{tildepropagator}
&[\mathbf{\widetilde {P}}_{k}^{[0]}]_{11}(p^2,t_1,t_2;\Phi_1,\Phi_2)=  \\&
G_k^+(p^2,\omega=0;\phi_1)[\Delta_k(\phi_{1},\phi_{2})-\widetilde R_k(p^2)]G_k^+(p^2,\omega=0;\phi_2)  
\end{aligned}
\end{equation}
and is thus purely static, up to a O($v$).

Difficulties however start to appear when one has to take derivatives of Eq. (\ref{gamma2_11}), {\it i.e.}, higher-order vertices built from the second cumulant. For instance we will need vertices like
\begin{equation}
\begin{aligned}
\label{gamma2_21}
&\frac{\delta}{\delta\hat\phi_{1,xt}}\bm{\mathsf\Gamma}_{k2;x_1t_1,x_2t_2}^{(11)}[\Phi_1, \Phi_2 ]\Big\vert_{\hat\phi_1= \hat\phi_2=0}= \\&
\delta^{(d)}(x,x_1,x_2) \delta(t-t_1)\Bigg( \begin{array}{cc}
0 \;&\; \Delta_k^{(10)}(\phi_{1,t_1},\phi_{2,t_2})\\
\\
0 \; &\; 0 \end{array} \Bigg)
\end{aligned}
\end{equation}
and 
\begin{equation}
\begin{aligned}
\label{gamma2_22}
&\frac{\delta^2}{\delta\hat\phi_{1,xt}\delta\hat\phi_{2,x't'}}\bm{\mathsf\Gamma}_{k2;x_1t_1,x_2t_2}^{(11)}[\Phi_1, \Phi_2 ]\Big\vert_{\hat\phi_1=\hat\phi_2=0}= \delta(t-t_1)\times   \\&
\delta(t'-t_2)\delta^{(d)}(x,x',x_1,x_2) \Bigg( \begin{array}{cc}
\Delta_k^{(11)}(\phi_{1,t},\phi_{2,t'}) \;&\; 0\\
\\
0 \; &\; 0 \end{array} \Bigg)
\end{aligned}
\end{equation}
where the fields are spatially uniform, $\delta^{(d)}(x,x_1,x_2)\equiv \delta^{(d)}(x-x_1)\delta^{(d)}(x_1-x_2)$ and similarly for $\delta^{(d)}(x,x',x_1,x_2)$.

If the second cumulant shows a cusp, $\Delta_k(\phi_1,\phi_2)\propto \vert \phi_1-\phi_2\vert$ when $\phi_1-\phi_2\to 0$, as one finds in relation with avalanches,\cite{narayan92,FRGledoussal-chauve,FRGledoussal-wiese,tarjus04,tissier11,tarjus13} the derivatives appearing in the above matrices are singular when the two arguments coincide. This problem, which is already encountered in the FRG treatment of an interface in a random environment,\cite{FRGledoussal-chauve} can be solved for the equilibrium and the hysteresis case, but in two distinct and specific ways. This is what we illustrate now.

\section{How to handle the cusp? An illustration}
\label{illustration}

We consider here the nonperturbative FRG flow equation for $J_k(\phi)$. After taking one derivative of Eq. (\ref{eq_flow_Gamma1}) with respect to $\hat\phi_{xt}$, using the definition in Eq. (\ref{eq_prescriptions}), replacing the $2\times 2$ matrices corresponding to the propagators and vertices in the chosen ansatz by the expressions given in section \ref{propagators}, and performing the trace, the flow can be diagrammatically expressed as
\begin{equation}
\begin{aligned}
\label{flow_Jk}
&\partial_k J_k(\phi_t)=\\&
\frac 12 \tilde{\partial}_k \int_q\int_{t_1t_2}
\Bigg\{\raisebox{-10pt}{\includegraphics[width=180pt,keepaspectratio]{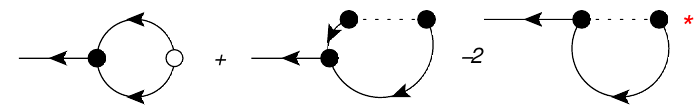}} \Bigg\}.
\end{aligned}
\end{equation}
where $\int_q \equiv \int d^dq/(2\pi)^d$. We have used the following graphical representation for the propagator (response function),
\begin{equation}
\label{gp}
G_k^+(q^2,t,t';\phi)=\raisebox{-10pt}{\includegraphics[width=60pt,keepaspectratio]{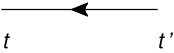}}
\end{equation}
and for the correlation function,
\begin{equation}
\label{gmp}
G_k(q^2,t,t';\phi)=\raisebox{-10pt}{\includegraphics[width=60pt,keepaspectratio]{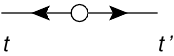}} \,,
\end{equation}
where an empty circle keeps track of $T X_k(\phi)$. In the propagators (and only in the propagators) the arrow indicates the direction of time: {\it i.e.}, in Eq. (\ref{gp}) the function is nonzero only if $t>t'$, whereas there is no ordering in Eq. (\ref{gmp}). In addition, a single filled circle denotes vertices obtained from the first cumulant $\mathsf{\Gamma}_{k1}$ and two filled circles joined by a dashed line denote vertices obtained from the second cumulant $\mathsf{\Gamma}_{k2}$. The legs of the vertices are associated with differentiation with respect to the fields: a leg with an incoming arrow indicates a derivative with respect to $\phi_{xt}$ and a leg with an outgoing arrow  a derivative with respect to $\hat\phi_{xt}$. So, for example,
\begin{equation}
\label{graph_delta}
\Delta_k(\phi_1,\phi_2)=\raisebox{0pt}{\includegraphics[width=70pt,keepaspectratio]{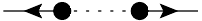}} \,.
\end{equation}

Of the three diagrams present in Eq. (\ref{flow_Jk}), the first one is exactly zero in $T=0$ and only the third one, marked with a red asterisk, potentially leads to an ambiguous expression  because it involves a first derivative of $\Delta_k$ evaluated for arguments corresponding to the same replica field (albeit considered at two different times). We therefore focus on its contribution to $\partial_k J_k(\phi)$ and consider the case of the ascending branch of the hysteresis curve. The contribution explicitly reads
\begin{equation}
\label{flow_Jk_anomalous}
-  \int_{t_2}\Delta_k^{(10)}(\phi_{t},\phi_{t_2})\tilde{\partial}_k  \int_q [G_k^+(q^2,t_2,t;\phi) +{\rm O}(v)]
\end{equation}
where we recall that for the present case $\phi_t=\phi+v t$ with $v\to 0^+$. Due to causality, $G_k^+(p^2,t,t_2;\phi)$ is zero for $t_2\geq t$. As a result, setting $t=0$ without loss of generality, one only needs to consider $\Delta_k^{(10)}(\phi,\phi +v t_2)$ when $v t_2 \to 0^-$, and Eq. (\ref{flow_Jk_anomalous}) can be rewritten as
\begin{equation}
\begin{aligned}
\label{flow_Jk_anomalous2}
&- \Delta_k^{(10)}(\phi,\phi^-)\tilde{\partial}_k  \int_q \int_{-\infty}^0 dt_2 G_k^+(q^2,t_2,0;\phi) +{\rm O}(v)\\&
= - \Delta_k^{(10)}(\phi,\phi^-)\tilde{\partial}_k  \int_q  \widehat P_k(q^2;\phi) +{\rm O}(v)
\end{aligned}
\end{equation}
where
\begin{equation}
\label{eq_hatP}
\widehat P_k(q^2;\phi)=\frac 1{J'_k(\phi) + Z_k(\phi)q^2+\widehat R_k(q^2)}
\end{equation}
is the static hat (or ``connected'') propagator and $\phi^-=\phi-0^+$.

To see that the above expression is unambiguous, it is useful to rewrite $\Delta_k(\phi_1,\phi_2)=\breve{\Delta}_k(\phi,\delta\phi)$, where $\phi=(\phi_1+\phi_2)/2$ and  $\delta\phi=(\phi_2-\phi_1)/2$. Then, $\Delta_k^{(10)}(\phi,\phi^-)= (1/2)[\breve{\Delta}^{(10)}_k(\phi,0)-\breve{\Delta}^{(01)}_k(\phi,0^-)]$. The first term gives a regular contribution, whether or not $\Delta_k$ has a cusp. The potentially anomalous contribution comes from the second term. $\breve{\Delta}_k(\phi,\delta\phi)$ is an even function of $\phi$ and $\delta\phi$ separately. So if it is smooth enough when $\delta\phi\to 0$, $\breve{\Delta}^{(01)}_k(\phi,0^\pm)=\breve{\Delta}^{(01)}_k(\phi,0)=0$. On the other hand, if it has a cusp,
\begin{equation}
\label{eq_cusp}
\breve{\Delta}_k(\phi,\delta\phi)= \breve{\Delta}_k(\phi,0)+\breve{\Delta}_{k,cusp}(\phi) \vert \delta\phi\vert + {\rm O}(\delta\phi^2),
\end{equation}
one finds that 
\begin{equation}
\label{eq_cusp_bis}
\breve{\Delta}^{(01)}_k(\phi,0^-)=-\breve{\Delta}^{(01)}_k(\phi,0^+)=-\breve{\Delta}_{k,cusp}(\phi)\,.
\end{equation}

One can now discuss more qualitatively the differences appearing in the equilibrium and the hysteresis situations when a cusp is present in $\Delta_k(\phi_1,\phi_2)$. In the hysteresis case, the cusp cannot be rounded because one has to work at $T=0$, but the ambiguity between the two sides of the cusp ({\it i.e.}, for $\delta\phi=0^\pm$) is lifted by the direction of the dynamics which is characterized by the sign of the vanishing velocity $v$. The ascending branch corresponds to one side of the cusp, when $v\to 0^+$, and the descending branch to the other side, when $v\to 0^-$. There is a global symmetry, vestige of the time-reversal and $Z_2$ symmetries, $\{\phi,v\} \to \{-\phi,-v\}$, between the two branches of the hysteresis curve but on each branch, both the time-reversal and the $Z_2$ symmetries are broken. This is quite different than the equilibrium case where both the time-reversal and $Z_2$ symmetries are satisfied (at least at the critical point in zero applied source). The ambiguity is then lifted by considering the limiting process by which $T\to 0$. Indeed, any nonzero temperature rounds the cusp and therefore guarantees that the derivative in the $\delta\phi$ direction is zero in $\delta\phi=0$:  $\breve{\Delta}^{(01)}_k(\phi,0)=0$ for $T>0$. The property is conserved when taking the limit $T\to 0$ via the thermal boundary layer mechanism\cite{balents-ledoussal,ledoussal10,tissier06,tissier11,balog_activated} and in some sense one should interpret $\breve{\Delta}^{(01)}_k(\phi,0)=0$ at $T=0$ as the symmetrized value $[\breve{\Delta}^{(01)}_k(\phi,0^+)+\breve{\Delta}^{(01)}_k(\phi,0^-)]/2=0$. (Finally, note that in Eq. (\ref{flow_Jk}) the contribution from the first diagram that involves the correlation function is irrelevant for both equilibrium and hysteresis criticality as it is equal to zero when $T=0$ while in the equilibrium case  with $T>0$, when cast in the appropriate dimensionless form, it flows to zero as one reaches the fixed point.\cite{tissier06,tissier11,balog_activated})

\section{Flow equations for the RFIM in and out of equilibrium}
\label{flow_eqs}

\subsection{Deriving the NP-FRG equations in the presence of a cusp}

By following the same line of reasoning as in the previous section, we have derived the RG flow equations for all the functions of our nonperturbative ansatz [Eqs. (\ref{eq_gammak1_ansatz}-\ref{eq_gammak3_ansatz})]. The details are given in Appendix \ref{app_flow} and we only display in what follows the potentially anomalous terms that are present in the hysteresis case but not in the equilibrium one. For each of the static functions, $J_k(\phi)$, $Z_k(\phi)$, $\Delta_k(\phi_1,\phi_2)$, we write
\begin{equation}
\begin{aligned}
\label{eqs_flow}
&\partial_s J_k(\phi) = \beta_{J,eq}(\phi)+\beta_{J,an}(\phi)\\&
\partial_s Z_k(\phi) = \beta_{Z,eq}(\phi)+\beta_{Z,an}(\phi)\\&
\partial_s \Delta_k(\phi_1,\phi_2) = \beta_{\Delta,eq}(\phi_1,\phi_2)+\beta_{\Delta,an}(\phi_1,\phi_2)
\end{aligned}
\end{equation}
where  we have introduced the dimensionless variable (``RG time'') $s=\ln(k/\Lambda)$. The expressions for the three $\beta_{eq}$ are those already derived for the equilibrium RFIM in the limit $T= 0$\cite{tarjus04,tissier11,balog_activated} and the anomalous contributions read 
\begin{equation}
\begin{aligned}
\label{eq_betaJ}
\beta_{J,an}(\phi)=\frac{1}{2}\breve{\Delta}_{k,cusp}(\phi)\int_q \partial_s\widehat R_k(q^2)\widehat P_k(q^2)^2
\end{aligned}
\end{equation}
\begin{equation}
\begin{aligned}
\label{eq_betaZ}
&\beta_{Z,an}(\phi)=\breve{\Delta}_{k,cusp}(\phi)\int_q \partial_s\widehat R_k(q^2)\widehat P_k(q^2)^3\Big\{J''_k(\phi)\widehat P_k(q^2)\times \\&
\Big[Z_k(\phi)+\widehat R'_k(q^2)+\frac {2q^2}{d}\Big (\widehat R''_k(q^2)-2[Z_k(\phi)+\widehat R'_k(q^2)]^2 \times \\&
\widehat P_k(q^2)\Big )\Big ] + Z'_k(\phi) \Big [-1+\frac{d+2}{d}q^2[Z_k(\phi)+\widehat R'_k(q^2)] \widehat P_k(q^2) \\&
+\frac{2q^4}{d} \widehat P_k(q^2)\Big (\widehat R''_k(q^2)-2[Z_k(\phi)+\widehat R'_k(q^2)]^2\widehat P_k(q^2)\Big )\Big ]\Big \}
\end{aligned}
\end{equation}
where primes denote derivation with respect to the argument and we have omitted for ease of notation the explicit $\phi$ dependence of $P_k(q^2;\phi)$. Finally, after switching from $\Delta_k(\phi_1,\phi_2)$ to $\breve{\Delta}_k(\phi,\delta\phi)$, one also has
\begin{equation}
\begin{aligned}
\label{eq_betaDelta}
&\beta_{\Delta,an}(\phi,\delta\phi)=\\& -\frac{1}{2}\Big\{\breve{\Delta}_{k,cusp}(\phi+\delta\phi)[\breve{\Delta}_{k}^{(10)}(\phi,\delta\phi)+\breve{\Delta}_{k}^{(01)}(\phi,\delta\phi)]\\&
\times \int_q \partial_s\widehat R_k(q^2)\widehat P_k(q^2;\phi+\delta\phi)^{3} + \breve{\Delta}_{k,cusp}(\phi-\delta\phi) \times \\&
[\breve{\Delta}_{k}^{(10)}(\phi,\delta\phi)-\breve{\Delta}_{k}^{(01)}(\phi,\delta\phi)]\int_q \partial_s\widehat R_k(q^2)\widehat P_k(q^2;\phi-\delta\phi)^{3} \Big \}\,.
\end{aligned}
\end{equation}
One can check that there is a symmetry between the ascending and the descending branches of the hysteresis loop which corresponds to inverting the fields {\it and} changing  $\breve{\Delta}_{k,cusp}$ in $-\breve{\Delta}_{k,cusp}$. Note also that, of course, all the anomalous contributions vanish in the absence of a cusp: the flow equations for the equilibrium and the out-of-equilibrium cases are then identical.

We have also obtained the flow equation for the kinetic function(s). (Note that these functions {\it do not} enter into the flow equations of the static functions $J_k$, $Z_k$, $\Delta_k$.) At equilibrium and for a nonzero temperature $T>0$, $X_k(\phi)=Y_k(\phi)$ as a result of the time-reversal symmetry and their flow has been studied in Ref. [\onlinecite{balog_activated}]. Temperature is irrelevant for the equilibrium critical behavior but is dangerously so. In particular, the thermal boundary layer mechanism discussed above leads to an activated dynamic scaling, which can be described at a phenomenological level in terms of droplet excitations. For the hysteresis case, $T=0$ and only $Y_k(\phi)$ has to be considered. Its flow cannot be written as the sum of a regular contribution equal to that of the equilibrium dynamics and an anomalous one because the equilibrium contribution is singular when $T=0$.\cite{balog_activated} Interestingly, the presence of an infinitesimal velocity $v$ now regularizes this otherwise singular behavior. One indeed finds
\begin{equation}
\begin{aligned}
\label{eq_flotY}
\partial_s Y_k(\phi) = &\beta_{Y,0}-\frac{1}{2}\breve{\Delta}_{k}^{(02)}(\phi,0^-)Y_k(\phi)\int_q \partial_s\widehat R_k(q^2)\widehat P_k(q^2)^3 \\&
+\beta_{Y,an}
\end{aligned}
\end{equation}
where $\breve{\Delta}_{k}^{(02)}(\phi,0^-)=\breve{\Delta}_{k}^{(02)}(\phi,0^+)$ is always a finite, regular function, contrary to $\breve{\Delta}_{k}^{(02)}(\phi,0)$ which diverges due to the presence of a $\delta(\delta\phi)$ function when there is a cusp. (In the equilibrium case this term indeed leads to a divergence in $T=0$, and relaxation dynamics must therefore be considered for $T>0$ where the divergence is rounded in a thermal boundary layer.\cite{balog_activated}). We give the explicit expressions of $\beta_{Y,0}$ and $\beta_{Y,an}$ in Appendix \ref{app_flow}.

\subsection{NP-FRG flow equations in a dimensionless form}

Finally, to cast the RG flow equations in a dimensionless form that allows one to investigate the critical physics at long length and time scales, one 
must introduce scaling dimensions. This is the second operation of any RG transformation in addition to the ``coarse-graining" leading to the dimensionful beta functions obtained above. 

The equilibrium critical point of the RFIM is controlled by a zero-temperature fixed point,\cite{villain84,fisher86} {\it i.e.}, a properly defined renormalized temperature $T_k$ flows to zero at the fixed point with an exponent $\theta>0$. This renormalized temperature can be defined by comparing the second and the first cumulants of the renormalized disorder, {\it e.g.}, $T_k \sim Z_k k^2/\Delta_k$. A subtlety here is that temperature also enters via the thermal noise in the Langevin equation: see Eq. (\ref{eq_stochastic_dynamics_RFIM}). The two temperatures are equivalent (and irrelevant) for the equilibrium criticality but this is not true for hysteresis: keeping a nonzero thermal noise destroys metastability and hysteresis, which is why we have to set this physical temperature to zero in order to study the out-of-equilibrium critical point of the driven RFIM. In this case we can nonetheless define a renormalized ``temperature", unrelated to any thermal bath, by the comparison between disorder cumulants, {\it i.e.}, $T_k \sim Z_k k^2/\Delta_k \sim k^\theta$. 

Near the critical point of the ascending branch of the hysteresis loop, one then has the following scaling dimensions:
\begin{equation}
\begin{aligned}
\label{eq_scaling_dimension}
&Z_{k} \sim k^{-\eta}, \; \phi-\phi_c  \sim k^{(d-4+\bar \eta)/2},\\&
J_k-J_c\sim k^{(d-2\eta+\bar \eta)/2}, \;\Delta_k \sim k^{-(2\eta- \bar \eta)},
\end{aligned}
\end{equation}
where $\phi_c$ and $J_c$ respectively denote the  values of the magnetization (field) and the magnetic field (source) at the 
out-of-equilibrium critical point, and, as in equilibrium, the exponents $\theta$, $\eta$ and $\bar\eta$ are related through $\theta=2+\eta-\bar\eta$.

Due to the lack of $Z_2$ inversion symmetry in the hysteresis case, two relevant parameters must be fine-tuned to reach the critical point. In practice, we 
account for the additional condition by defining a displaced field variable $\phi-\phi_{c,k}$ where $\phi_{c,k}$ is fixed such that the second derivative of the renormalized force is zero all along the flow, $J''_k(\phi_{c,k})=0$, which corresponds to the field (magnetization) for which the connected susceptibility is maximum. When approaching the fixed point, $\phi_{c,k}\to \phi_c + k^{(d-4+\bar \eta)/2} \varphi_0$ and similarly, $J(\phi_{c,k})\to J_c + k^{(d-2\eta+\bar \eta)/2} j_0$. If the critical system flows to a fixed point where $Z_2$ symmetry is restored, then $\varphi_0=j_0=0$ but otherwise they are different from zero and depend on the initial conditions.

Using lower-case letters, $j'_k, z_k,\delta _k, \varphi, \delta\varphi$, to denote the dimensionless counterparts of 
$J'_k, Z_k,\Delta _k, \phi, \delta\phi$,  the dimensionless form of the flow equations can be symbolically written as
\begin{equation}
\label{eq_flow_dimensionless}
\begin{aligned}
&\partial_s j'_k(\varphi)= \\&
-(2-\eta_k)j'_k(\varphi)  +\frac 12(d-4+\bar\eta_k)\varphi j''_k(\varphi)+ \beta_{j',k}(\varphi),\\&
\partial_s z_k(\varphi)= \eta_k + \frac 12(d-4+\bar\eta_k)\varphi z'_k(\varphi) + \beta_{z,k}(\varphi),\\&
\partial_s \breve\delta_k(\varphi,\delta\varphi)=\\&
 (2\eta_k-\bar\eta_k)\breve\delta_k(\varphi,\delta\varphi) +\frac 12(d-4+\bar\eta_k)(\varphi \partial_ \varphi +\\&  \delta\varphi\partial_{\delta\varphi})\breve\delta_k(\varphi,\delta\varphi) + \beta_{\delta,k}(\varphi,\delta\varphi)\, ,
\end{aligned}
\end{equation}
where the now dimensionless beta functions in the right-hand sides themselves depend on $j'_k$, $z_k$, $\breve\delta_k$ and their derivatives. They also depend on the dimensionless IR cutoff function $r(q^2/k^2)=\widehat R_k(q^2)/(Z_k k^2)$. [The other IR cutoff function is also related to $r$ via $\widetilde R_k(q^2)= - \Delta_k r'(q^2/k^2)$.] The running anomalous dimensions $\eta_k$ and $\bar\eta_k$ are fixed by the condition that $z_k(\varphi_{0,z})=1$ and $\breve\delta_k(\varphi_{0,\delta},0)=1$, respectively, where $\varphi_{0,z}$ and $\varphi_{0,\delta}$ are appropriately chosen points (close to the $\varphi_0$ introduced before but not necessarily equal). Note that we have considered for further convenience the flow of the derivative $j'_k(\varphi)$ in place of that of $j_k(\varphi)$, but the set of coupled flow equations formed by Eq.~(\ref{eq_flow_dimensionless}) is still closed.

Finally, in the hysteresis case, the kinetic coefficient $Y_k(\phi)\sim k^{2-\eta -z} y_k(\varphi)$, where $z$ is the dynamical exponent characterizing the relation between duration and size of the critical avalanches, satisfies the following RG flow equation in dimensionless form:
\begin{equation}
\label{eq_flow_dimensionless}
\begin{aligned}
&\partial_s y_k(\varphi)= \\&
-(2-\eta_k -z_k) y_k(\varphi)  +\frac 12(d-4+\bar\eta_k)\varphi y'_k(\varphi)+ \beta_{y,k}(\varphi),\\&
\end{aligned}
\end{equation}
where the dynamical exponent $z_*$ at the fixed point is obtained from an eigenvalue equation, once the other functions and anomalous dimensions have been determined.

\section{Results}
\label{results}

We have numerically solved the set of coupled flow equations discussed in the previous section both by (i) directly studying the fixed-point equations (root finding method), and then investigating the stability of the solutions,  and by (ii) running the flow from initial conditions and searching for the fixed point(s) by dichotomy. This can be done as a function of dimension $d$, starting from the Gaussian fixed point at the upper critical dimension $d=6$. The numerical resolution requires discretizing the fields on a fine grid and is rather involved. In the equilibrium case, we obtain solution down to $d\approx 2.7$ whereas for the hysteresis case, numerical problems start to hamper the resolution for $d\approx 3.5$ and below. The numerical methods and the associated technical difficulties are discussed in detail in Appendix \ref{app_numerics}.

\subsection{$d>d_{DR}$}

\begin{figure}[ht]
\begin{center}
\includegraphics[width=\linewidth]{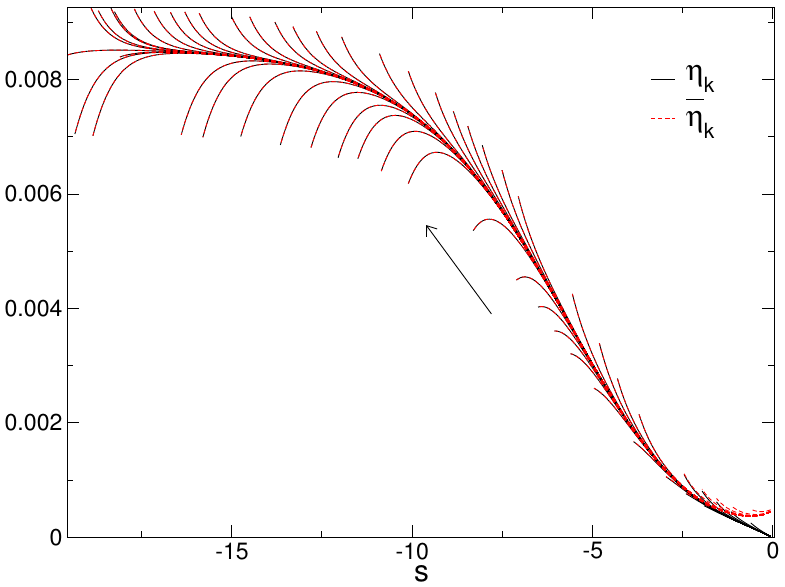}
\caption{Critical behavior on the ascending branch of the hysteresis curve in $d=5.5$: RG flows of the anomalous dimensions $\eta_k$ and $\bar\eta_k$ starting from a variety of initial non-$Z_2$ cuspy conditions. By dichotomy one can fine-tune the initial conditions to reach the fixed point as $s=\ln(k/\Lambda) \to -\infty$. Except for a short transient regime the two anomalous dimensions become indistinguishable (and are then equal to their equilibrium counterparts). The arrow indicates the direction of the RG flow.}
\label{Fig_4}
\end{center}
\end{figure}

The first important result is that the fixed points controlling the in and out of equilibrium critical points are the same above the critical dimension $d_{DR}\approx 5.1$ separating the region where the fixed-point dimensionless cumulants have no cusp and the $d\to (d-2)$ dimensional reduction property applies ($d>d_{DR}$) from that where the fixed-point dimensionless cumulants display a cusp and dimensional reduction is not valid ($d<d_{DR}$). Hysteresis and equilibrium criticality are in the same universality class for $d \geq 5.1$: The linearized RG equations around the fixed point lead to identical critical exponents (which are given by the dimensional-reduction property).\cite{footnote_cuspy_eigenvalue} In this range of dimensions, this confirms the conclusion of our previous study:\cite{balog-tissier-tarjus} Despite the fact that the driven system is out of equilibrium and that there is no $Z_2$ symmetry at the hysteresis critical points, the $Z_2$ symmetry is restored along the RG flow and the fixed point coincides with the equilibrium one. This results from the property that the cusp in the renormalized second cumulant of the random field, which stems from the presence of avalanches in the zero-temperature evolution of the system, is subdominant at long distance. The various anomalous contributions to the beta functions therefore lead to subdominant behavior and do not change the fixed point (nor the scaling behavior). 

\begin{figure}[ht]
\begin{center}
\includegraphics[width=0.98\linewidth]{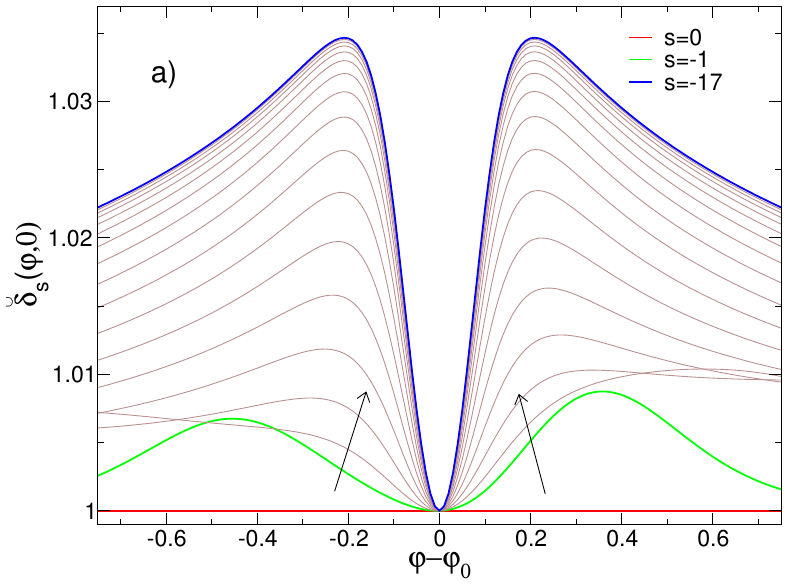}
\includegraphics[width=\linewidth]{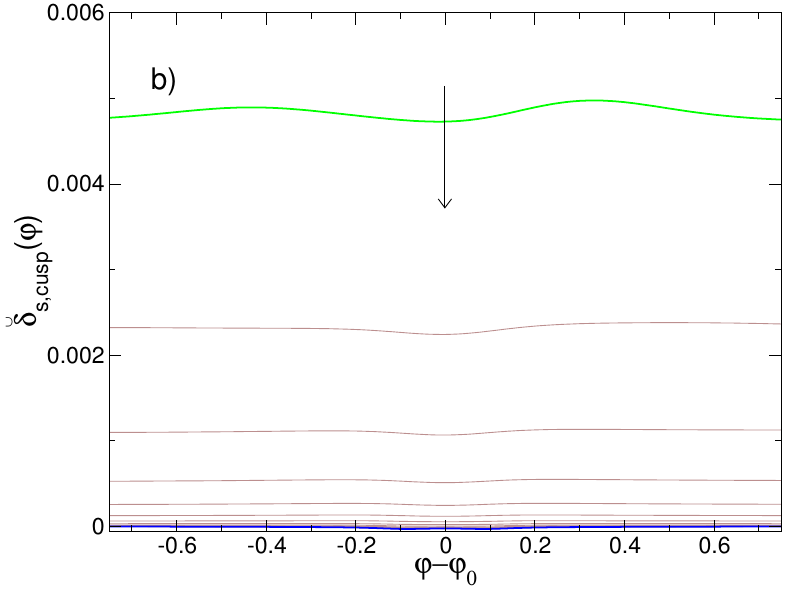}
\caption{Critical behavior on the ascending branch of the hysteresis curve in $d=5.5$: (a) Dimensionless second cumulant $\breve\delta_k(\varphi,0)$ for a variety of values of the RG time $s=\ln(k/\Lambda)$ and initial non-$Z_2$ cuspy conditions that are fine-tuned to lead to the fixed point. (b) Same for the  cusp amplitude $\breve\delta_{k,cusp}(\varphi)$. The initial cusp amplitude was put to a constant value of $1$ and is not shown for practical reasons. The green curve is for $s=-1$ and the amplitude has already dropped by a factor of $200$. The arrows indicate the direction of the RG flow.}
\label{Fig_5}
\end{center}
\end{figure}

This is illustrated in Figs.~\ref{Fig_4} and \ref{Fig_5} for $d=5.5$. In Fig.~\ref{Fig_4} we display the flows of the anomalous dimensions $\eta_k$ and $\bar\eta_k$  for a variety of initial non-$Z_2$ conditions. We have included a cusp for the initial conditions on $\breve\delta_k(\varphi,\delta\varphi)$ (as would result from a mean-field description of avalanches). Fig.~\ref{Fig_4} illustrates the procedure of dichotomy that allows us to reach the fixed point.  One can see that the two exponents $\eta_k$ and $\bar\eta_k$ rapidly become indistinguishable as the flow progresses and are strictly equal at the fixed point, which is a signature of the dimensional-reduction property. They are also equal then to their equilibrium counterparts (and to the value obtained for the pure $\phi^4$ theory in dimension $d-2$).
In Fig.~\ref{Fig_5} we display the evolution with RG time of the functions $\breve\delta_k(\varphi,0)$ and $\breve\delta_{k,cusp}(\varphi)$ when starting from initial non-$Z_2$ cuspy conditions that are fine-tuned to lead to the fixed point. As the flow progresses the dimensionless second cumulant $\breve\delta_k(\varphi,0)\equiv \delta_k(\varphi,\varphi)$ develops an asymmetry around the value $\varphi_0$ at which it has a minimum but becomes asymptotically symmetric again [see Fig.~\ref{Fig_5}(a)]: this corresponds to a restoration of the $Z_2$ symmetry at the fixed point. The function is then identical to its equilibrium counterpart. Furthermore, as shown in Fig.~\ref{Fig_5}(b), the amplitude of the cusp goes to zero asymptotically: the cusp is indeed irrelevant at the fixed point. The rapid decrease of $\breve\delta_{k,cusp}(\varphi)$ at the beginning of the flow explains the rapid convergence of the two anomalous dimensions seen in Fig.~\ref{Fig_4}.

\subsection{$d<d_{DR}$}

\begin{figure}[ht!]
\begin{center}
\includegraphics[width=\linewidth]{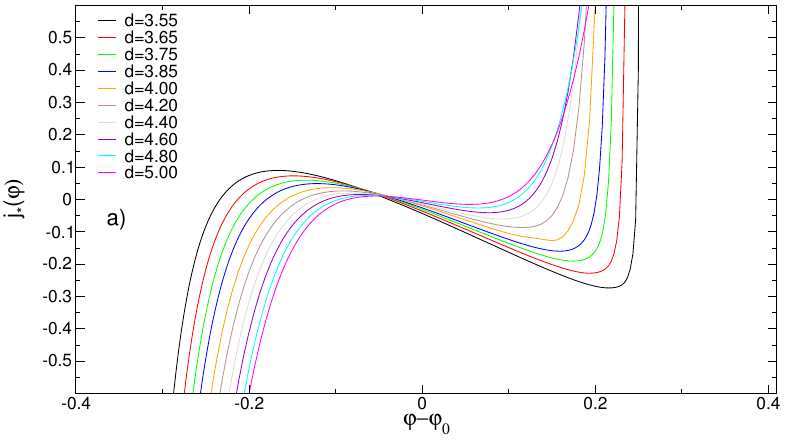}
\includegraphics[width=\linewidth]{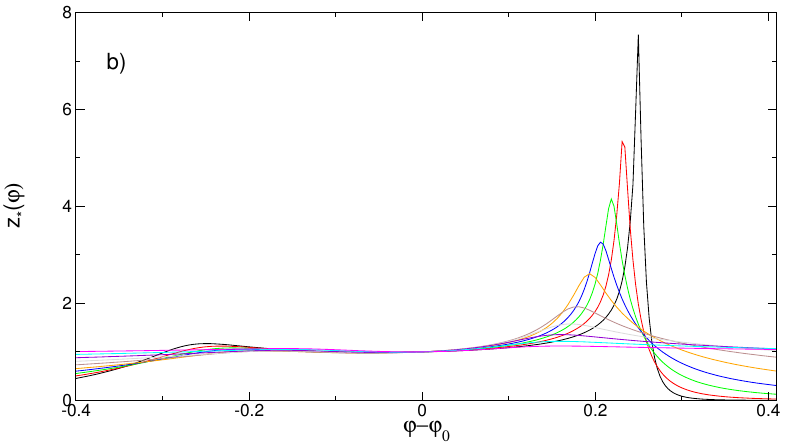}
\includegraphics[width=\linewidth]{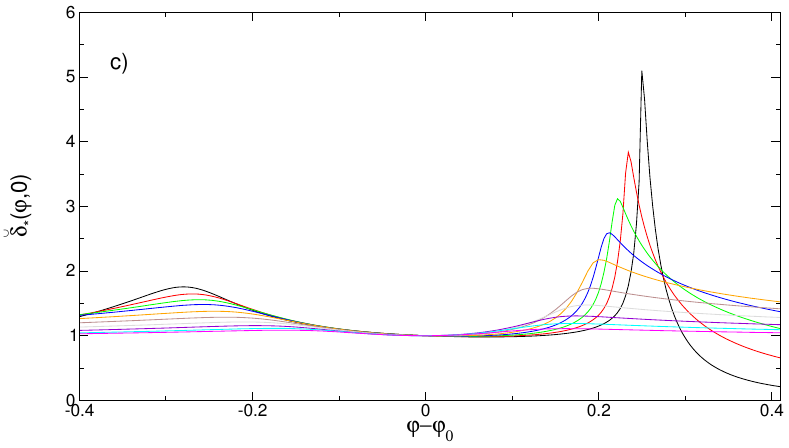}
\includegraphics[width=\linewidth]{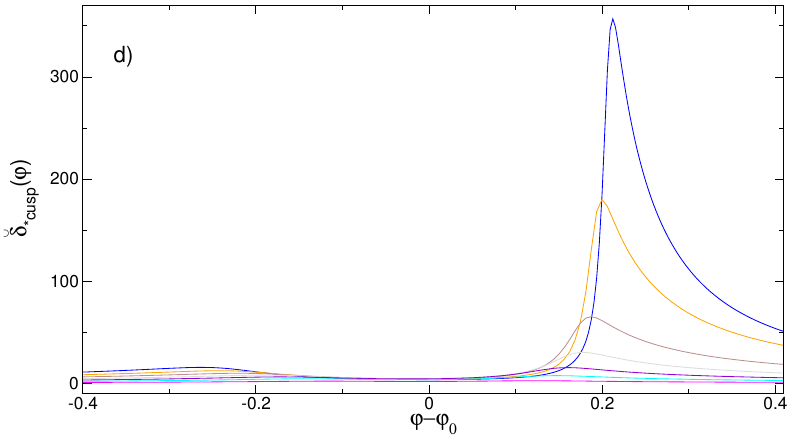}
\caption{Critical behavior on the ascending branch of the hysteresis curve for $d<d_{DR}$: Fixed-point functions $j_*(\varphi)$, $z_*(\varphi)$, $\breve\delta_*(\varphi,0)$, and  $\breve\delta_{*,cusp}(\varphi)$  plotted versus $\varphi-\varphi_0$ for several values of $d$ between $5$ and $3.5$.}
\label{Fig_6}
\end{center}
\end{figure}

When considering dimensions lower than $d_{DR}$ we find however that the fixed point for the equilibrium and that for hysteresis differ. Both are cuspy ({\it i.e.}, $\breve\delta_{*,cusp}(\varphi)\neq 0$) but the anomalous contributions to the beta functions coming from the cusp produce a difference. In consequence, the equilibrium and out-of-equilibrium critical behaviors of the RFIM are {\it not} in the same universality class for $d \lesssim 5.1$, which includes the physical dimension $d=3$. This is the phenomenon that we have previously missed by working with the static superfield formalism.\cite{balog-tissier-tarjus} 

When looking at the fixed-point functions, the difference between equilibrium and hysteresis becomes striking as one lowers the dimension. We display in Fig. \ref{Fig_6} the functions $j_*(\varphi)$, $z_*(\varphi)$, $\breve\delta_*(\varphi,0)$, and  $\breve\delta_{*,cusp}(\varphi)$ characterizing the hysteresis critical fixed points for several values of $d$ between $5$ and $3.5$. (They are plotted versus $\varphi-\varphi_0$, where $\varphi_0$ is defined in the previous section.) $Z_2$ symmetry at the fixed point would correspond to $j_*(\varphi)$ antisymmetric with respect to $\varphi-\varphi_0$ and all the other functions symmetric with respect to $\varphi-\varphi_0$. This is approximately realized in $d=5$ but becomes increasingly violated as $d$ is lowered.

\begin{figure}[ht]
\begin{center}
\includegraphics[width=\linewidth]{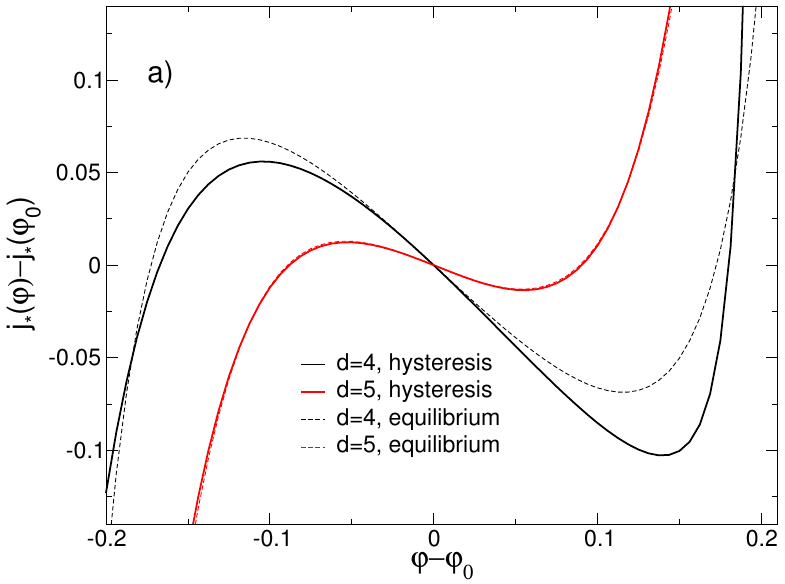}
\includegraphics[width=\linewidth]{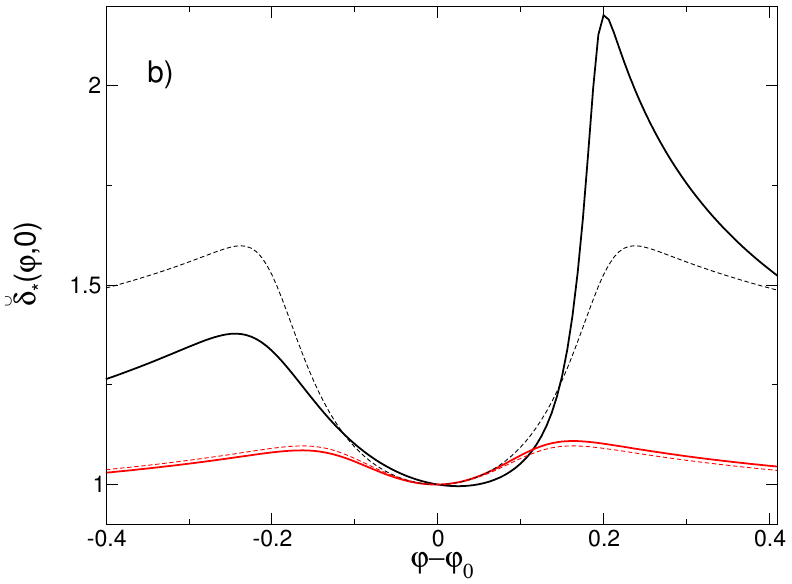}
\caption{Comparison of the equilibrium and hysteresis fixed-point functions $j_*(\varphi)$ (a) and $\breve\delta_*(\varphi,0)$ (b) for $d=4$ and $d=5$ (versus $\varphi-\varphi_0$ with $\varphi_0=0$ and $j_*(\varphi_0)=0$ for equilibrium).}
\label{Fig_7}
\end{center}
\end{figure}

In Fig. \ref{Fig_7} we compare two of the fixed-point functions,  $j_*(\varphi)-j_*(\varphi_0)$ and $\breve\delta_*(\varphi,0)$, which are obtained for the equilibrium and hysteresis critical points in $d=4$ and $d=5$. While the difference is small and barely noticeable for $d=5$ it becomes conspicuous in $d=4$.

It is worth trying to relate these fixed-point functions to scaling functions that could be measured in simulations (or experiments). This is what we have already discussed in section \ref{summary} summarizing the results. Unfortunately, the functions that  more dramatically display the absence of $Z_2$ symmetry in the hysteresis case, hence the difference with the equilibrium functions, are $z_*(\varphi)$, $\breve\delta_*(\varphi,0)$, and $\breve\delta_{*,cusp}(\varphi)$, which are hard to access in simulations. More readily accessible functions that can be extracted from finite-size scaling are reduced susceptibilities, which are also obtainable from the fixed-point functions. The scaling functions associated with the connected and disconnected susceptibilities (see  section \ref{summary}) are indeed expressed as
\begin{equation}
\begin{aligned}
&\hat f(\varphi)=\frac 1{j'_*(\varphi) +r(0)} \\&
\tilde f(\varphi)= \frac{\breve\delta_{*}(\varphi,0)}{[j'_*(\varphi) +r(0)]^2} \,.
\end{aligned}
\end{equation}
As illustrated in Fig. \ref{Fig_FSS}, the asymmetry and the difference with equilibrium found for the hysteresis case are not as striking as found for the functions plotted in Figs. \ref{Fig_FSSratio}, \ref{Fig_6} and \ref{Fig_7}. Although this could be in part obscured by the uncertainty in the finite-size scaling procedures, a careful comparison should allow a crisper conclusion than the previously done comparison of integrated avalanche functions and critical exponents.\cite{maritan94,perez04,liu09}

\subsection{Critical exponents}

\begin{figure}[ht]
\begin{center}
\includegraphics[width=\linewidth]{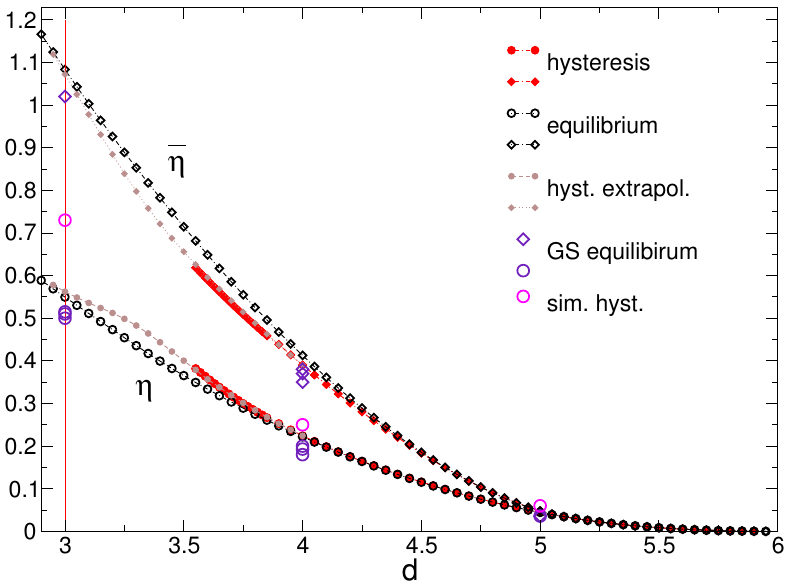}
\caption{Anomalous dimensions $\eta$ and $\bar\eta$ versus $d$ for the equilibrium and hysteresis critical fixed points. This also includes the extrapolated points for hysteresis obtained from the $\lambda$-modified FRG equations as described in the text. The symbols represent the results of  ground-state computations for equilibrium\cite{hartmann,middleton,fytas} and computer simulations for hysteresis.\cite{perkovic} (The error bars are not shown but one should note that for hysteresis the uncertainty on $\eta$ is very large.\cite{perkovic})}
\label{Fig_8}
\end{center}
\end{figure}

Despite the striking difference in the equilibrium and hysteresis fixed-point functions for $d < 5$, the difference in the anomalous dimensions associated with these fixed points appears very small. This is shown in Fig. \ref{Fig_8} where we plot $\eta$ and $\bar\eta$ as a function of $d$ for equilibrium and hysteresis. In both cases $\eta$ and $\bar\eta$ start to separate when $d<d_{DR}$, as a result of the cuspy nature of the fixed points and of the breakdown of dimensional reduction (the exponent $\theta=2-\eta-\bar\eta$ become less than $2$). However, the equilibrium and hysteresis values stay within $5\%$. 

\begin{figure}[ht]
\begin{center}
\includegraphics[width=\linewidth]{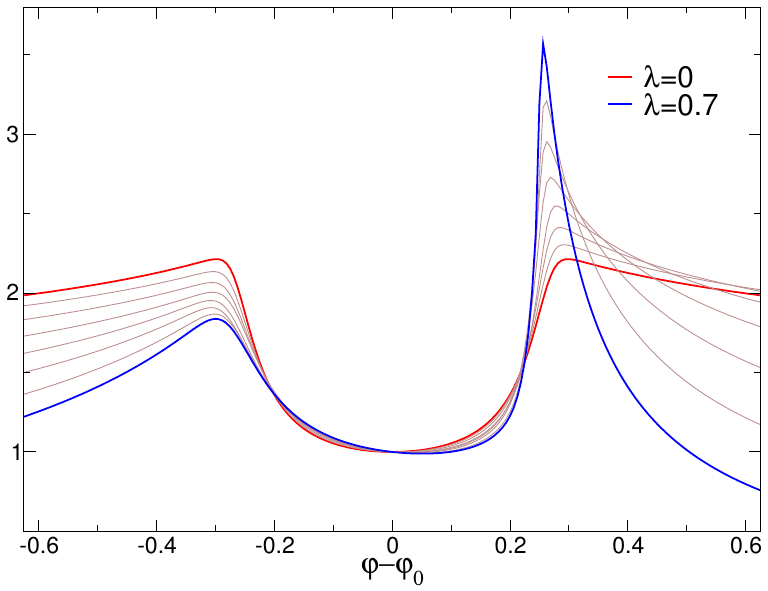}
\caption{Solution of the $\lambda$-modified FRG equations (see text)  in $d=3.55$: Fixed-point function $\breve\delta_*(\varphi,0\vert \lambda)$ for several values of the parameter $\lambda \in [0,1]$.}
\label{Fig_9}
\end{center}
\end{figure}

The dashed lines in Fig. \ref{Fig_8} are hysteresis data obtained for $d\lesssim 3.5$ through an extrapolation procedure. We have studied modified RG flow equations where we have added a multiplying factor $\lambda \in [0,1]$ in front of the anomalous contributions in all of the beta functions. This is a purely technical trick that gives the equilibrium result when $\lambda=0$ and the hysteresis one when $\lambda=1$. For $d<3.52$, we encounter numerical difficulties to solve the fixed point equations when $\lambda$ is close to $1$, and more so as $d$ decreases. Fig. \ref{Fig_9} illustrates the problem for the function $\breve\delta_*(\varphi,0\vert \lambda)$ in $d=3.55$: as $\lambda$ increases the right peak in the function becomes increasingly sharp and at some point it becomes impracticable to use a grid that is fine enough to resolve the peak and large enough to capture the behavior at large fields. More details are given in Appendix \ref{app_numerics}. The exponents gathered for a range of values of $d\leq4$ and $\lambda$ are shown in this Appendix, and various extrapolation procedures are used to obtain estimated values for the hysteresis case ($\lambda=1$) down to $d=3$. The extrapolated values are robust and are shown as the dashed lines in Fig. \ref{Fig_8}.

\begin{figure}[ht]
\begin{center}
\includegraphics[width=\linewidth]{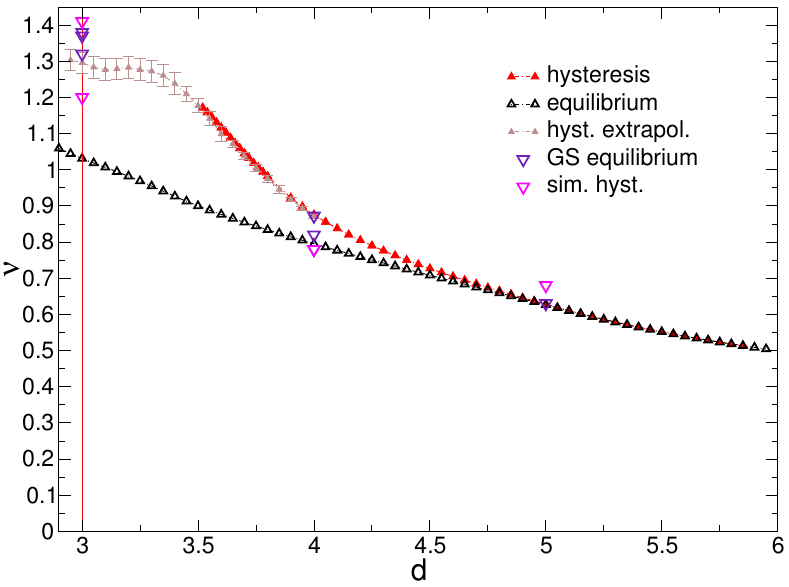}
\caption{Correlation length exponent $\nu$ versus $d$ for equilibrium and hysteresis critical points. This also includes the extrapolated points for hysteresis obtained from the $\lambda$-modified FRG equations as described in the text. The symbols represent the results of  ground-state computations for equilibrium\cite{hartmann,middleton,fytas} and computer simulations for hysteresis.\cite{perkovic,perez03} (The error bars are not shown but one should note that the uncertainty on the value of $\nu$ for hysteresis, which varies between $1.2$\cite{perez03} and $1.4$\cite{perkovic} in $d=3$, is very large.)}
\label{Fig_10}
\end{center}
\end{figure}

The data for the correlation length exponent $\nu$ are displayed in Fig. \ref{Fig_10}. Here the difference between hysteresis and equilibrium values is more significant and reaches about $25\%$. Note however that the NP-FRG prediction (with the present truncation) for the exponent $\nu$ at equilibrium does not compare as well with recent large-scale simulation data in $d=3$\cite{hartmann,middleton,fytas} as those for the anomalous dimensions $\eta$, $\bar\eta$ which are properties of the fixed point. The difference found between equilibrium and hysteresis is of the order of the discrepancy observed for the equilibrium exponent in $d=3$. Note also that the numerical values obtained for the hysteresis critical behavior from computer simulation for the exponent $\nu$ in $d=3$ have a large dispersion, between $1.2 \pm0.1$\cite{perez03} and $1.4\pm0.1$,\cite{perkovic} and are compatible with the value that we predict.

\begin{figure}[ht]
\begin{center}
\includegraphics[width=\linewidth]{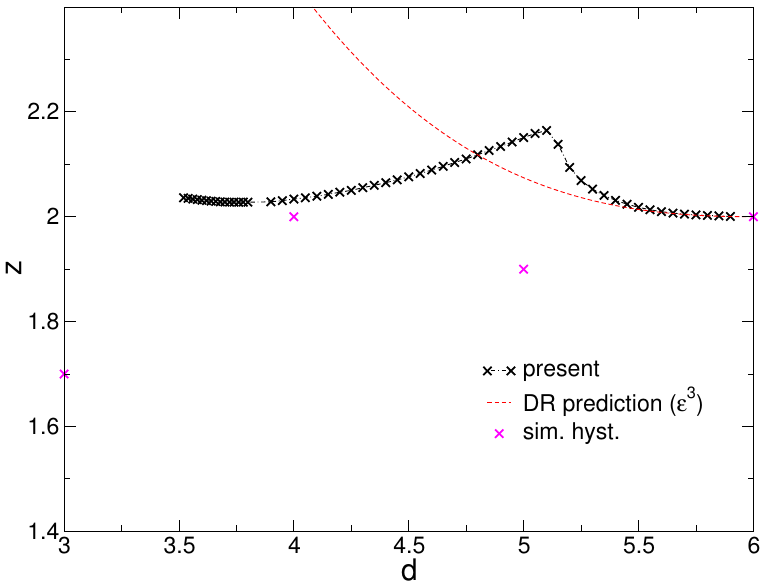}
\caption{Dynamical critical exponent $z$  versus $d$ for the hysteresis critical point(s). The symbols represent the results of a computer simulation\cite{perkovic} (the error bars are not shown but the uncertainty is quite large). Note that close to $d=6$ when dimensional reduction applies, $z$ should be given by the $\epsilon$ expansion of the pure model: the result\cite{krey} obtained at order O($\epsilon^3$) is shown as the red dashed line.}
\label{Fig_11}
\end{center}
\end{figure}

Finally, we present the results for the dynamical exponent $z$ in Fig. \ref{Fig_11}. One sees that is stays close to $2$, its mean-field value, over the whole range of dimension, as also found in computer simulations.\cite{perkovic} Here there is no comparison with equilibrium because the critical slowing down for relaxation to equilibrium is described by activated dynamic scaling (for $T>0$) and there is no analog of the exponent $z$ in this case (in a sense $z$ is infinite).

\section{Conclusion}
\label{conclusion}

We have shown that contrary to expectations based on numerical simulations\cite{maritan94,perez04,liu09} or erroneous theoretical treatment,\cite{footnote_perturbativeRG,balog-tissier-tarjus} the critical points of the RFIM in equilibrium and out of equilibrium (hysteresis curve) are {\it not} in the same universality class in low enough dimension, {\it i.e.}, below $d\sim 5$. The difference in critical behavior comes from the presence of avalanches in the evolution of the system at zero temperature under a change of the applied source, avalanches which have been shown to be relevant for the long-distance behavior only below a critical dimension $d_{DR}\approx 5.1$.

As the exponents are close for hysteresis and equilibrium critical behaviors and may be within numerical or experimental uncertainty, we have stressed that  the proper way to check whether the two types of critical points belong (or not) to the same universality class is to consider reduced (dimensionless) functions that are properties of the critical fixed points and are sensitive to the presence (or not) of $Z_2$ inversion symmetry. The integrated distribution of avalanche sizes and related functions that have been previously studied in computer simulations are unable to probe this, and we have proposed several candidates that can be accessed through finite-size studies.

Finally, we point out that the fact that equilibrium and hysteresis critical points are controlled by different fixed points in $d<d_{DR}\approx 5.1$ does not imply that the lower critical dimension is different in the two cases. While $d=2$ has been rigorously proven as being the lower critical dimension for the equilibrium RFIM,\cite{aizenman} it has been suggested on the basis of numerical studies\cite{vives_2D,spasojevic_2D,shukla_2D} that the hysteresis critical behavior of the RFIM could have a lower critical dimension strictly lower than $2$. (This latter conclusion has however been challenged by Sethna and coworkers by means of an extended scaling derived from normal form theory.\cite{raju17,sethna_private}) Unfortunately, the NP-FRG approach with the present approximation scheme is not well suited to address this issue. As illustrated in this work, it is indeed technically hard to solve the RG equations below $d=3$ (or even $d\approx 3.5$ for hysteresis). The lower critical dimension for systems with an Ising symmetry is not directly nor exactly accessible through truncations of the spatial derivative expansion even in the pure case\cite{ballhausen,rulquin} and it is unclear at present how to improve the approximation scheme to resolve this problem.

\begin{acknowledgments}
IB acknowledges the support of the Croatian Science Foundation Project No. IP-2016-6-7258 and of the QuantixLie center of excellence.
\end{acknowledgments}

\appendix

\section{Derivation of the NP-FRG flow equations for the RFIM in and out of equilibrium}
\label{app_flow}

In this appendix we describe in detail how to derive the flow equations for hysteresis and equilibrium and we comment on the difference between these two cases. As the derivation for the equilibrium case and the role of temperature there have already been discussed in previous publications,\cite{tarjus04,tissier06,tissier11,balog_activated} we put the emphasis on the hysteresis behavior, more precisely on the ascending branch of the hysteresis curve (when $v\to 0^+$).

\subsection{Flow of $\Delta_k$} 

We start by discussing the flow of the second cumulant of the renormalized random field, $\Delta_k(\phi_1,\phi_2)$. After taking two derivatives of Eq. (\ref{eq_flow_Gamma2}) with respect to $\hat\phi_{1,xt}$ and $\hat\phi_{2,x't'}$, using the definition in Eq. (\ref{eq_prescriptions}), replacing the $2\times 2$ matrices corresponding to the propagators and vertices in the chosen ansatz by the expressions given in section \ref{propagators}, and performing the trace, the flow equation can be diagrammatically expressed as
\begin{equation}
\begin{aligned}
\label{flowdelt}
&\partial_s\Delta_k(\phi_{1},\phi_{2}) =\\& 
\frac 12 \tilde{\partial}_s  \int_q\int_{t_1t_2}
\Bigg\{\raisebox{-45pt}{\includegraphics[width=170pt,keepaspectratio]{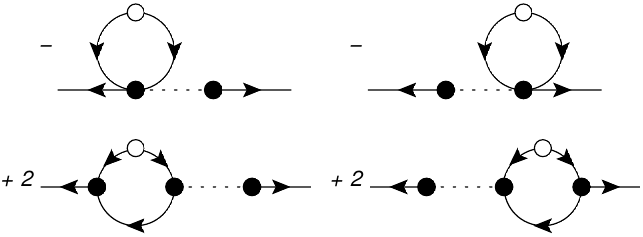}} \\&
\raisebox{-70pt}{\includegraphics[width=220pt,keepaspectratio]{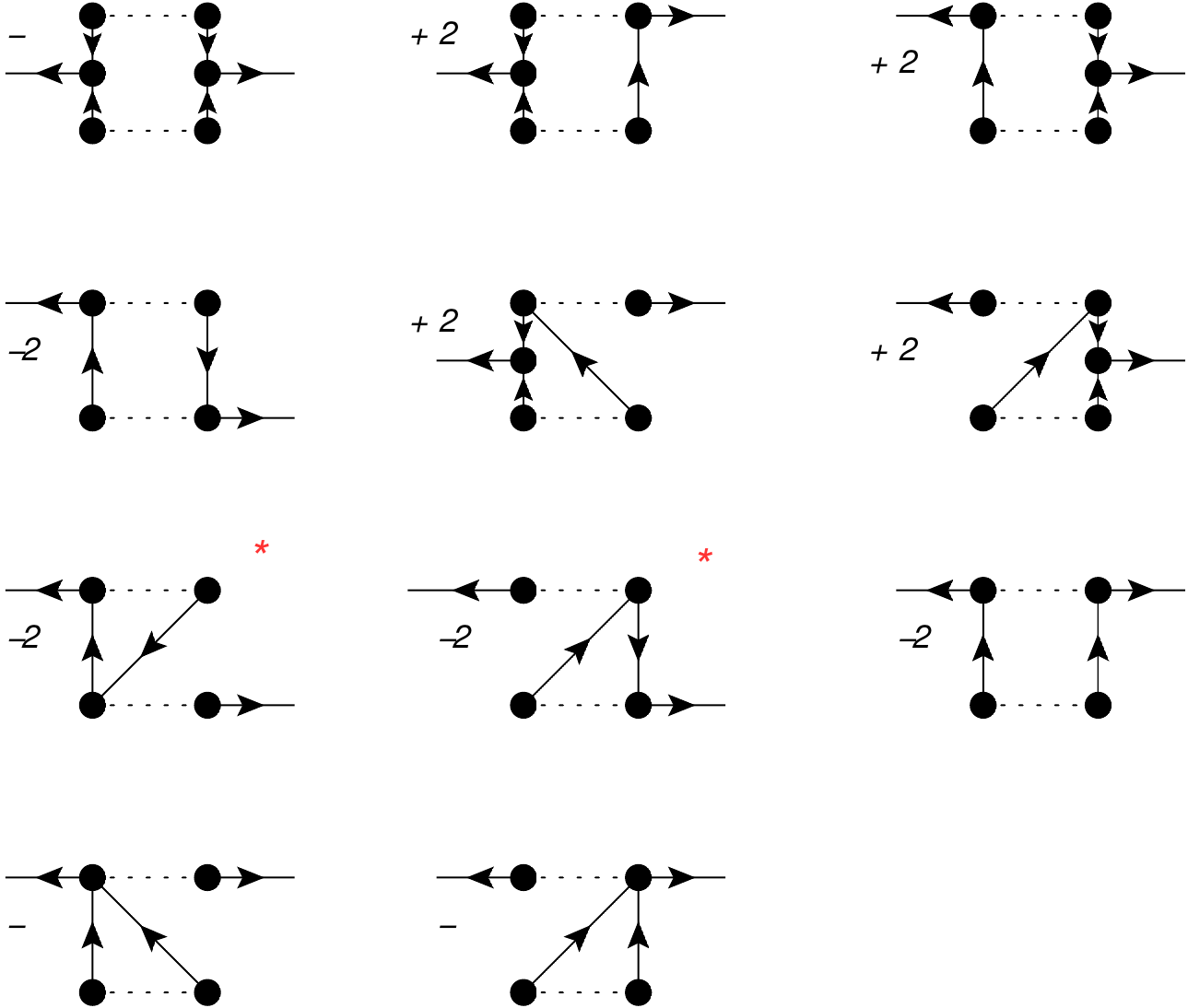}} \raisebox{-60pt} {\Bigg\}.}
\end{aligned}
\end{equation}
The graphical representation has been explained below Eq. (\ref{flow_Jk}) of the main text. In all diagrams the left external leg corresponds to the replica field $\phi_1$ and the right one to the field $\phi_2$.

The first 4 diagrams are exactly zero for $T=0$. In the equilibrium case and for $T>0$ the two first ones give rise to a term proportional to  $T[\Delta^{(20)}_k(\phi_1,\phi_2)+\Delta^{(02)}_k(\phi_1,\phi_2)]$, which after changing variables to $\phi=(\phi_1+\phi_2)/2$ and $\delta\phi=(\phi_2-\phi_1)/2$ can be reexpressed as $(T/2)[\breve\Delta_k^{(02)}(\phi,\delta\phi)+\breve\Delta_k^{(20)}(\phi,\delta\phi)]$. The second derivative in the $\delta\phi$ direction makes it impossible for a cusp to develop near $\delta\phi=0$ when $T>0$ (otherwise the derivative would blow up). This is the origin of the rounding of the cusp in a thermal boundary layer. However, when rescaling all quantities to look for a fixed point, the contributions of all of the four diagrams flow to zero.\cite{tissier06,tissier11,balog_activated} We can therefore always drop them when studying directly the fixed point, for hysteresis as well as for equilibrium.

The other $11$ diagrams survive at $T=0$. One can easily check that they are all well defined in the presence of a cusp in $\Delta_k(\phi_1,\phi_2)\equiv \breve\Delta_k^{(02)}(\phi,\delta\phi)$, except possibly the two diagrams marked with a red asterisk. The contribution of these two diagrams can be explicitly written as
\begin{equation}
\begin{aligned}
\label{flow_delta_dangerous}
&\tilde{\partial}_s  \int_q\int_{t_1t_2}\Bigg\{\Delta^{(10)}_k(\phi_2(t_2),\phi_1(0))\Delta^{(01)}_k(\phi_2(t_1),\phi_2(0))\\&
\times G_k^+(q^2,t_2-t_1;\phi_2)G_k^+(q^2,-t_2;\phi_2)+perm(12)\Bigg\}.
\end{aligned}
\end{equation}
where we have arbitrarily chosen $t=t'=0$ for the external legs. The term that is potentially ill-defined is $\Delta^{(01)}_k(\phi_2(t_1),\phi_2(0))$ and the one obtained by permutation between $\phi_2$ and $\phi_1$. To handle this for the equilibrium case we keep an infinitesimal temperature $T$ and take field configurations that are constant in time. One can then integrate over the times $t_1$ and $t_2$, which provides the static (hat) propagator $\widehat P_k(q^2)$ defined in Eq. (\ref{eq_hatP}), and we use the property that when the cusp is rounded, $\Delta^{(10)}_k(\phi,\phi)+\Delta^{(01)}_k(\phi,\phi)=\breve\Delta^{(01)}_k(\phi,\delta\phi=0)=0$ by symmetry. We finally obtain the contribution in the form
\begin{equation}
\begin{aligned}
\label{flow_delta_dangerous_eq}
&\frac{1}{4}\tilde{\partial}_s  \int_q \Big\{\breve{\Delta}^{(10)}_{k}(\phi+\delta\phi,0)[\breve{\Delta}_{k}^{(10)}(\phi,\delta\phi)+\breve{\Delta}_{k}^{(01)}(\phi,\delta\phi)]\\& \times
 \widehat P_k(q^2;\phi+\delta\phi)^2 + \breve{\Delta}^{(10)}_{k}(\phi-\delta\phi,0) [\breve{\Delta}_{k}^{(10)}(\phi,\delta\phi)- \\&
\breve{\Delta}_{k}^{(01)}(\phi,\delta\phi)] \widehat P_k(q^2;\phi-\delta\phi)^2 \Big \}\,,
\end{aligned}
\end{equation}
which is now well-behaved when $T\to 0$.

One has to proceed differently for the hysteresis case at $T=0$. As already explained in the main text we now consider an infinitesimal velocity $v\to 0^+$  with the fields taken as $\phi_a(t)=\phi_a +v t$. An ordering of the times $t_1$ and $t_2$ in Eq. (\ref{flow_delta_dangerous}) is provided by the property of the response function and It\=o's prescription. For the first term in the curly brackets, one must have $0>t_2>t_1$; this term then becomes in the limit $v\to 0^+$ and after integration over the two times, $\tilde{\partial}_s  \int_q \Delta^{(10)}_k(\phi_2,\phi_1)\Delta^{(01)}_k(\phi_2,\phi_2^+)\widehat P_k(q^2;\phi_2)^2$, where $\phi_2^+=\phi_2+0^+$. The potential ambiguity arising from the cusp is lifted because 
\begin{equation}
\begin{aligned}
\Delta^{(01)}_k(\phi_2,\phi_2^+)&=\frac 12 \breve\Delta^{(10)}_k(\phi+\delta\phi,0) +\frac 12 \breve\Delta^{(01)}_k(\phi+\delta\phi,0^+) \\&
=\frac 12 \breve\Delta^{(10)}_k(\phi+\delta\phi,0) +\frac 12 \breve\Delta_{k,cusp}(\phi+\delta\phi) \,,
\end{aligned}
\end{equation}
where we have used Eqs. (\ref{eq_cusp}) and (\ref{eq_cusp_bis}). A similar result is derived for the term obtained by permutation, so that the contribution of the two diagrams to the flow of $\Delta_k$ in the hysteresis case is equal to the contribution at equilibrium given above plus
\begin{equation}
\begin{aligned}
&\frac{1}{4}\tilde{\partial}_s  \int_q\Big\{\breve{\Delta}_{k,cusp}(\phi+\delta\phi)[\breve{\Delta}_{k}^{(10)}(\phi,\delta\phi)+\breve{\Delta}_{k}^{(01)}(\phi,\delta\phi)]\\&
\times \widehat P_k(q^2;\phi+\delta\phi)^{2} + \breve{\Delta}_{k,cusp}(\phi-\delta\phi) 
[\breve{\Delta}_{k}^{(10)}(\phi,\delta\phi)- \\&
\breve{\Delta}_{k}^{(01)}(\phi,\delta\phi)]\widehat P_k(q^2;\phi-\delta\phi)^{2} \Big \}\,.
\end{aligned}
\end{equation}
When using the definition of the operator $\tilde{\partial}_s$ this leads to Eq. (\ref{eq_betaDelta}).

\subsection{Flows of the 1-copy functions}

To derive the flow equations for the functions $J'_k$ (which corresponds at equilibrium to the second derivative of the effective potential $U_k''$), $Z_k$ and $Y_k$ one starts from the common equation,
\begin{equation}
\begin{aligned}
\label{flowd2G1}
 &\partial_s\frac{\delta^2\mathsf{\Gamma}_{k1}[\Phi]}{\delta\phi_{p t}\delta\hat{\phi}_{p't'}} = (2\pi)^d \delta^{(d)}(p+p') \, \frac 12 \tilde{\partial}_s {\rm tr} \int_q\int_{t_1t_2} \\& 
 \Bigg\{\raisebox{-10pt}{\includegraphics[width=220pt,keepaspectratio]{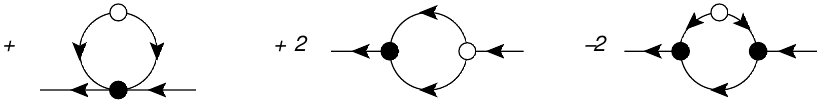}} \\&
 \raisebox{-70pt}{\includegraphics[width=240pt,keepaspectratio]{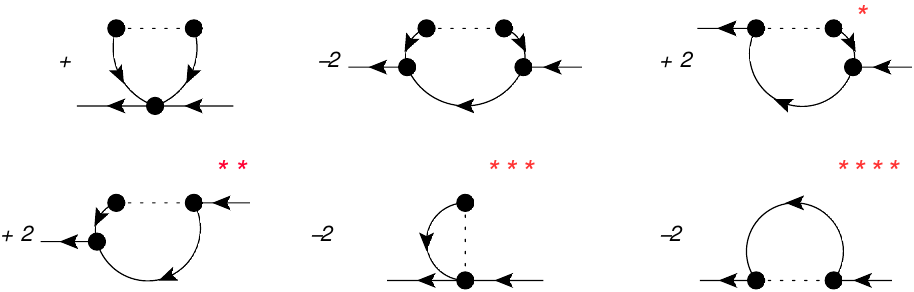}}\raisebox{-60pt}{\Bigg\}\,,}
\end{aligned}
\end{equation}
 giving the flow of the second functional derivative of the first cumulant with respect to the fields $\hat{\phi}$ and $\phi$ for the same copy, but for nonzero external momenta and different times. The configuration of the fields needed to obtain the flows of $J'_k(\phi)$, $Z_k$ and $Y_k$ [see Eq. (\ref{eq_prescriptions})] are spatially uniform and either constant in time (equilibrium) or such that $\phi(t)=\phi +v t$ in the limit $v \to 0^+$ (ascending branch of the hysteresis).

The three diagrams of the first line of Eq. (\ref{flowd2G1}) are zero at $T=0$ and can be dropped when studying the fixed point for the same reasons as discussed above. Of the 6 remaining diagrams, which all survive in $T=0$, only the last 4 (all marked with red asterisks) are potentially dangerous in the presence of a cusp in the second cumulant $\Delta_k$ because they involve derivatives of this cumulant that are evaluated when the two arguments correspond to the same replica (copy) field. Since the flow of $J'_k(\phi)$ can also be obtained by deriving once the flow of $J_k(\phi)$ given in the main text, we will focus on $Z_k(\phi)$ and $Y_k(\phi)$.

We consider first the flow of the (static) field renormalization function $Z_k(\phi)$. It is easy to realize that due to the constraints on the momenta appearing in the internal loop the last two diagrams (marked with 3 and 4 red asterisks) do not have any dependence on the external momentum $p$ and therefore do not contribute to the flow of $Z_k$. Since we are studying a static function, we can take the external time $t'=0$ and integrate over the other external time $t$ to obtain a zero-frequency quantity.

The contribution of the diagram marked by two red asterisks can then be explicitly expressed as
\begin{equation}
\begin{aligned}
\label{eq_Ztwo}
&\tilde{\partial}_s  \int_q\int_0^\infty dt_2\int_{-\infty}^{t_2}dt_1 [J''_k(\phi)+Z'_k(\phi)(q^2+p^2-q p)] \times \\&
\Delta^{(10)}_k(\phi,\phi+v t_1) G_k^+(q^2,t_2-t_1;\phi)G_k^+(\vert p-q\vert^2,t_2;\phi)\,,
\end{aligned}
\end{equation}
where we have used the properties of the response function to restrict the ranges of the time integrals. We have also kept for now an infinitesimal velocity $v$ [Eq. (\ref{eq_Ztwo}) is valid up to a O($v$)] and considered directly $T=0$. By taking into account the expression of the response function in eq. (\ref{response_function}), splitting the integration over $t_1$ into a contribution from $-\infty$ to $0$ (for which $\phi+v t_1\to \phi^-$) and one from $0$ to $t_2$ (for which $\phi+v t_1\to \phi^+$), and introducing the variables $\phi$ and $\delta\phi$, one can rewrite the above expression as
\begin{equation}
\begin{aligned}
&\frac 12 \breve\Delta^{(10)}_k(\phi,0) \,\tilde{\partial}_s  \int_q  \, \widehat P_k(q^2;\phi)\widehat P_k(\vert p-q\vert ^2;\phi) \\& \times [J''_k(\phi)+Z'_k(\phi)(q^2+p^2-q p)] \,.
\end{aligned}
\end{equation}
Its contribution to the flow of $Z_k$  is obtained by deriving with respect to $p^2$ and evaluating the derivative in $p^2=0$. It is therefore not anomalous.
 
The only anomalous contribution to the flow of $Z_k$ comes from the diagram marked by a single red asterisk. Proceeding along the same lines as before, one finds that its contribution to Eq. (\ref{flowd2G1}) in zero frequency can be written in the limit $v\to 0^+$ as
\begin{equation}
\begin{aligned}
&\frac 12[\breve\Delta^{(10)}_k(\phi,0) + \breve\Delta_{k,cusp}(\phi,0)]\tilde{\partial}_s  \int_q \widehat P_k(q^2;\phi)\widehat P_k(\vert p-q\vert ^2;\phi) \\& \times [J''_k(\phi)+Z'_k(\phi)(q^2+p^2-q p)] 
\,.
\end{aligned}
\end{equation}
After taking a derivative with respect to $p^2$ and evaluating it in $p^2=0$, the anomalous contribution proportional to $\breve\Delta_{k,cusp}(\phi,0)$ is equal to that given in Eq. (\ref{eq_betaZ}). The regular contribution is part of the equilibrium beta function whose complete expression is given elsewhere.\cite{tissier11} [Note that at equilibrium it is the sum of the above two diagrams marked with one and two asterisks that is well-defined in the limit $T\to 0$ because one then uses the symmetry-related property that $\Delta^{(10)}_k(\phi,\phi)+\Delta^{(01)}_k(\phi,\phi)=\breve\Delta^{(10)}_k(\phi,0)$.]

Finally, we derive the flow of the kinetic function $Y_k(\phi)$. We take the same field configurations as before, zero external momentum ($p=0$), and we Fourier transform over time to be able to extract the term then proportional to $-i\omega$ (this should be done in the quasi-static limit $v\to 0^+$ for hysteresis). The contribution of the diagram marked with 3 red asterisks is then zero. 

Let us first consider the contribution of the diagram marked with 4 red asterisks. It gives a term proportional to $-i\omega Y_k(\phi)$ that is equal to 
\begin{equation}
\begin{aligned}
&\Delta^{(11)}_k(\phi,\phi^-) \tilde{\partial}_s  \int_q  \widehat P_k(q^2;\phi)^2 \,,
\end{aligned}
\end{equation}
in the hysteresis case. The second derivative of the second cumulant, $\Delta^{(11)}_k$, can be rewritten as $(1/4)[\breve\Delta^{(20)}_k(\phi,0)+\breve\Delta^{(02)}_k(\phi,0^-)]$ and is well-behaved even in the presence of a cusp since the second derivative $\breve\Delta^{(02)}_k(\phi,0^-)=\breve\Delta^{(02)}_k(\phi,0^+)$ is evaluated on a given side of the cusp and not strictly in $\delta\phi=0$ (the second derivative $\breve\Delta^{(20)}_k$ is always finite and unambiguous). This corresponds to the second term in the right-hand side of Eq. (\ref{eq_flotY}). For equilibrium on the other hand one has to consider $\Delta^{(11)}_k(\phi,\phi)$ with strictly equal arguments. In the presence of a cusp this derivative blows up and the flow is ill-defined. One should instead take $T>0$: the rounding of the cusp in a thermal boundary layer guarantees that $\Delta^{(11)}_k(\phi,\phi) \propto 1/T$ and gives rise to activated dynamic scaling.\cite{balog_activated}

The two diagrams marked with 1 and 2 red asterisks are regular in the equilibrium case, even at $T=0$ in the present of a cusp, but lead to additional anomalous contributions in the hysteresis case. We do not detail here the derivation which follows the lines already described and we give the final result for this anomalous contribution to the flow of $Y_k(\phi)$:
\begin{equation}
\begin{aligned}
\beta_{Y,an}=&\frac{1}{2}\breve{\Delta}_{k,cusp}(\phi) \int_q \partial_s\hat{R}_k(q^2)\hat{P}_k(q^2;\phi)^3\big \{2 Y'_k(\phi) - \\&
\frac{3}{2}\hat{P}_k(q^2;\phi) Y_k(\phi) [J'_k(\phi) + Z'_k(\phi) q^2]\big\}\,.
\end{aligned}
\end{equation}
Finally, for the sake of completeness, we provide the regular piece of the flow, denoted $\beta_{Y,0}$ in Eq. (\ref{eq_flotY}):
\begin{equation}
\begin{aligned}
 &\beta_{Y,0}= \frac{1}{4}\int_{q}\Bigg\{ -2\partial_s\tilde{R}_k(q^2) \hat{P}_k(q^2;\phi)^2 Y_k''(\phi) 
- 4 \hat{P}_k(q^2;\phi)^3 \\& 
 \times [J'_k(\phi) + Z_k'(\phi) q^2 ] \Big(-2 \big [\partial_s\tilde{R}_k(q^2) + 3 \partial_s\hat{R}_k(q^2) \hat{P}_k(q^2;\phi) \\&
 \times (\breve{\Delta}_k(\phi,0)-\tilde{R}_k(q^2) )\big ] Y_k'(\phi)  - 3 \partial_s\hat{R}_k(q^2)  \hat{P}_k(q^2;\phi) \times \\&
 \breve{\Delta}_k^{(10)}(\phi,0) Y_k(\phi) \Big) - 4 \hat{P}_k(q^2;\phi)^4  Y_k(\phi)\big [\partial_s\tilde{R}_k(q^2) + \\&4
 \partial_s\hat{R}_k(q^2) (\breve{\Delta}_k(\phi,0)-\tilde{R}_k(q^2)) \hat{P}_k(q^2;\phi) \big ]\big [J'_k(\phi) + Z_k'(\phi)q^2  \big ]^2 \\&
 - 2 \partial_s\hat{R}_k(q^2)\hat{P}_k(q^2;\phi)^3 \Big[2 (\breve{\Delta}_k(\phi,0)-\tilde{R}_k(q^2)) Y_k''(\phi) + \\& 
4\breve{\Delta}_k^{(10)}(\phi,0)  Y_k'(\phi) + \breve{\Delta}_k^{(20)}(\phi, 0)Y_k(\phi) \Big ] \Bigg\} \,.
\end{aligned}
\end{equation}

\section{Numerical resolution and related difficulties}
\label{app_numerics}

\subsection{Numerical resolution}

Solving the NP-FRG equations requires extensive numerical calculations, for either following the RG flows (which we have done only for $d\geq d_{DR}\approx 5.1$) or directly searching for the fixed-point solutions. In the present approximation, the calculation involves 3 functions $j'_k(\varphi)$, $z_k(\varphi)$ and $\breve\delta_k(\varphi,\delta\varphi)\equiv \delta_k(\varphi_1=\varphi-\delta\varphi, \varphi_2=\varphi+\delta\varphi)$ characterizing the statics of the system and one function $y_k(\varphi)$ for the dynamics. The latter can be treated separately as it does not enter into the set of coupled RG equations for the 3 static functions.

Most of the data presented in this paper have been obtained from fixed-point solutions of the system. To search for such solutions we have implemented the Newton-Raphson root finding procedure using Fortran 90. All the RG functions are discretized on a grid (details later in the text) for some initial condition. Then perturbing each discretized value by a small amount ($\approx 10^{-8}$) allows us to determine the matrix of the derivatives of the flow, {\it i.e.}, the stability matrix $M$. We then invert the stability matrix to obtain from the initial condition via the Newton-Raphson step the RG functions for the next iteration. We keep iterating until the fixed point is reached to a satisfying precision. The latter is assessed through a ``normalized flow" that is defined as 
\begin{equation}
D=\sqrt{\sum_i\frac{\int_{\varphi}\beta_{f_i}(\varphi)^2}{\int_{\varphi}f_i(\varphi)^2}}\, ,
\end{equation} 
where $f_i$ denote the set of RG functions and $\beta_{f_i}$ their associated beta function. We typically obtain fixed points with a precision up to $D=10^{-13}$. Once we have the fixed point solution, we can diagonalize the stability matrix to obtain all the eigenvalues characterizing this fixed point. Depending on the discretization mesh, we obtain in practice reliable results for the lowest 3 or 4 eigenvalues. Two of them correspond to relevant directions (they are negative, since the RG time $s=\ln(k/\Lambda)$ goes to $-\infty$), {\it i.e.}, $-1/\nu$ and the exponent characterizing the $\varphi^3$ perturbation; the third one is the exponent $\omega$ controlling the dominant corrections to scaling.  
 
To compute the RG flows going into the vicinity of the fixed point, we have used the Euler method, also implemented in Fortran 90. (We have found no benefit from using the 4th order Runge-Kutta method over the simpler Euler method for calculating the flows since precision data have been obtained from the alternative procedure based on root finding and stability analysis of the fixed points.) To approach the fixed point along the RG flow, we have used a procedure based on ``dichotomy". We first pick a range of initial values for the ``mass" $j'(\varphi_0)$ (which is a relevant control parameter) and for each value we run the flow. If the flow goes into the paramagnetic phase, we then restrict the range of masses by taking this initial mass value as an upper bound. Conversely, if the flow goes into the ferromagnetic phase, we then restrict the range of masses by taking this initial mass value as a lower bound. This procedure is efficient for finding a flow that goes very close to the fixed point: see {\it e.g.} Fig. \ref{Fig_4} where we can follow the RG flow to a ``time" $s\sim -18$, {\it i.e.}, down to an IR cutoff $k\sim 10^{-7}\Lambda$, and get very close to the fixed point. There is an additional complication when following the RG flows for the hysteresis case compared to the equilibrium one, which arises from the lack of the $Z_2$ symmetry. Since the $\varphi^3$ coupling is always a  relevant perturbation, to run the dichotomy such that one reaches the close vicinity of the critical fixed point, one additional parameter has to be fine-tuned besides the mass: the location of the minimum of $j_k'$, {\it i.e.}, $\varphi_{0,k}$. In practice we solve this problem (for $d\geq d_{DR}$) by using the input from the root finding and we define the grid so that $\varphi_{0,k}$ is always at the center point along the RG flow. This is achieved by constraining $j_k''(\varphi_{0,k})=0$ at all RG times.    

For convenience we solve the static problem by working with $j'(\varphi)$ in place of $j(\varphi)$. The FRG equations are still closed for the three functions $j'(\varphi)$, $z(\varphi)$ and $\breve\delta(\varphi,\delta\varphi)$.  In this way the expressions of the beta functions only require derivatives of order up to 2 of the functions we are determining. This minimizes numerical errors (if we had instead  worked with the function $j$ itself, we would have needed derivatives up to order 3 of this function). We discretize the field dependences on $\varphi$ and $\delta\varphi$ on a square grid. The field $\varphi$ is considered within the domain $[-\varphi_M+\varphi_0,\varphi_M+\varphi_0]$ and is discretized on $2n_x+1$ points in this range. Since the nontrivial domain of the solution is not necessarily centered around $\varphi=0$ as a result of the lack of $Z_2$ symmetry, we find it convenient to control the offset field $\varphi_0$ as a function of the dimension $d$ to better capture the solution with least numerical effort (see below). 

The second cumulant $\delta$ is a function of 2 fields $\varphi$ and $\delta\varphi$ and it is discretized on a trapezoidal domain such that for a given field, say $\delta\varphi_1$, the needed range of field $\varphi$ is within the domain $[-(\varphi_M-\delta\varphi_1)+\varphi_0,(\varphi_M-\delta\varphi_1)+\varphi_0]$. The reason is that the beta function for, {\it e.g.}, $\breve\delta(\varphi_1,\delta\varphi_1)$ requires RG functions at $\delta\varphi=0$ and $\varphi=\varphi_1 \pm\delta\varphi_1$: it is then clear that if the function $\breve\delta$ is defined on such a trapezoidal domain, all the data required to determine its flow inside the domain is contained within it. The field $\delta\varphi$ is discretized on $n_y+1$ points between $0$ and $\delta\varphi_M$. We have used finite-difference formulas through 5 points to determine the higher derivatives that are  required in the beta functions.  

The chosen boundaries $\varphi_M$ and $\delta\varphi_M$ change with dimension $d$ and have been determined in a way not to influence the results up to at least 5 relevant digits. In principle $\varphi_M$ should be large enough for the nontrivial contribution to the beta functions to be negligible in comparison to the trivial dimensional contribution, and the same should also apply at the corners of the trapezoid for the largest $\delta\varphi$ (where $\varphi=\varphi_0\pm(\varphi_M-\delta\varphi_M)$ and $\delta\varphi=\delta\varphi_M$).  When the nontrivial contribution to the beta functions is negligible, and the dimensional contribution dominates, the solution is then very close to a power-law dependence. The power-law exponent is positive for the function $j'$, hence this function goes to $\infty$ for large fields, and is negative for $z$ and $\delta$, so that these functions go to $0$. (The large field dependence of the function $\breve\delta$ on  $\delta\varphi$ is the same as for the field $\varphi$.)

We have used the knowledge of the large field asymptotic behavior in $\varphi$ to implement a trick that greatly stabilizes the numerical procedures: We  have extended the grid by a certain number of points that we ``predict" based on the power-law dependence and the values near the edges. This has enabled us to use bulk formulas for the finite-difference derivatives on the large $|\varphi+\varphi_0|$ edges of the grid. We suspect that the reason why this trick stabilizes the numerics is that it somehow mimics imposing boundary conditions at large fields. One knows that imposing such asymptotic boundary conditions together with the requirement that the functions be defined and continuous over the whole range guarantees the uniqueness of the relevant fixed points.\cite{RGreview} (We found that imposing a similar ``boundary condition" on the edges of the function $\breve\delta$ for $\delta\varphi=\delta\varphi_M$ and $\delta\varphi=0$ was not necessary and we therefore did not enforce it for simplicity.) The trick is especially beneficial for dimensions  $d<4.5$.

Using a range of $\varphi$ as large as possible is useful for numerical stability because it decreases the nontrivial part of the beta functions at the edges. However, there is a trade-off since the RG functions possess a structure and one needs to use a fine enough grid to correctly capture all the features. Using large fine grids obviously slows down the computation considerably. For illustration, our best results, going as low as $d=3.5$ for hysteresis, has been calculated on a grid with $n_x=220$ and $n_y=70$, $\varphi_M=0.6875$ and $\delta\varphi_M=0.21875$. For dimensions larger than $d=3.8$ we have observed that we can use $n_x=110$ and $n_y=35$ with the same range of extreme field values. Using the larger grid becomes  numerically very demanding since our root finding implementation draws $\approx 20GB$ of RAM and finding one fixed point lasts approximately $1$ day, even with parallelization. For comparison, for the smaller of the two lattices, the program draws $2GB$ of RAM and finding one fixed point lasts approximately $30$ min.    

We have mentioned before that because of the lack of $Z_2$ symmetry it is important to control the offset of the grid in the field $\varphi$. It appears that the nontrivial region of the solution shifts as one lowers $d$. We have implemented the root finding in a way that the center of the grid is always approximately in the middle between the peaks of the $z$ function, which is close to the field $\varphi_0$ already introduced.

\subsection{Technical difficulties}

For $d<4.5$ in the hysteresis case, there appears one technical difficulty, which is due to the large value of the function $j'(\varphi)$ at the edges of the grid. This function is positive for large $|\varphi|$ and for the mentioned dimensions it can reach $\sim 10^8$. With the Newton-Raphson method where one constructs the stability matrix $M$ by perturbing every value of the RG functions by a small amount, typically $\sim 10^{-8}$, it is obvious that when the value of the function itself reaches $10^8$ there is no precision left in the 16 digits of accuracy contained in the ``double" real number for the perturbation to make a difference. To solve this problem for $d<4.5$ we have considered in place of  $j'(\varphi)$ the function 
\begin{equation}
e(\varphi)=\ln[j'(\varphi)+r(0)]
\end{equation} 
which stays within reasonable bounds for all available dimensions (it reaches  at worst $\approx 70$ at the positive edge of the grid of $\varphi$ for $d\approx 3.5$). Reformulating all the flow equations in terms of this function is a straightforward task.  We have also checked that for $d>4.5$, using the function $e(\varphi)$ or $j'(\varphi)$ makes no difference for the critical exponents up to 4 relevant digits.  

A more serious problem arises as one decreases the dimension: We can find fixed points down to only  $d=3.5$ in the hysteresis case while with the same kind of precision and with an almost identical numerical procedure (up to moving the grid) we can reach down to $d=2.7$ for the equilibrium case. The reason why the numerical procedure fails is because it runs out of precision to describe the functions $z(\varphi)$ and $\delta(\varphi, \varphi)=\breve\delta(\varphi,0)$ in the vicinity of their peaks. 

We see as an empirical fact that the variation of the functions $z$ and $\breve\delta$ around their sharpest peak ({\it e.g.}, the right one when considering the ascending branch of the hysteresis: see Figs. \ref{Fig_6}, \ref{Fig_7}, \ref{Fig_9}) is much stronger for hysteresis than for equilibrium at the same dimension.  We have found that the development of such strong features as $d$ decreases is associated with the behavior of several eigenvalues of the stability matrix $M$, the second and the third one, whose absolute values become quite small ($<0.05$). Thanks to a singular value decomposition of the stability matrix we have been able to find that these eigenvalues are sensitive to how well the solution is captured around the steep peak. When the precision is insufficient, these eigenvalues can fluctuate close to $0$, making the stability matrix near singular and hence noninvertible. This is the reason why the Newton-Raphson method for root finding fails. 

Note that this failure is unrelated to a too close approach to the lower critical dimension, as we are always quite far from the latter. Indeed, near the lower critical dimension one would expect the value of $j'(\varphi_0)$ to get close to $-r(0)$ (so that a pole is approached in the dimensionless propagator at zero momentum). Yet we observe that the value is nowhere near this limit in either equilibrium or hysteresis at the lowest dimension $d$ that can be reached.

\subsection{$\lambda$-modified FRG equations}

\begin{figure}
\begin{center}
\includegraphics[width=\linewidth]{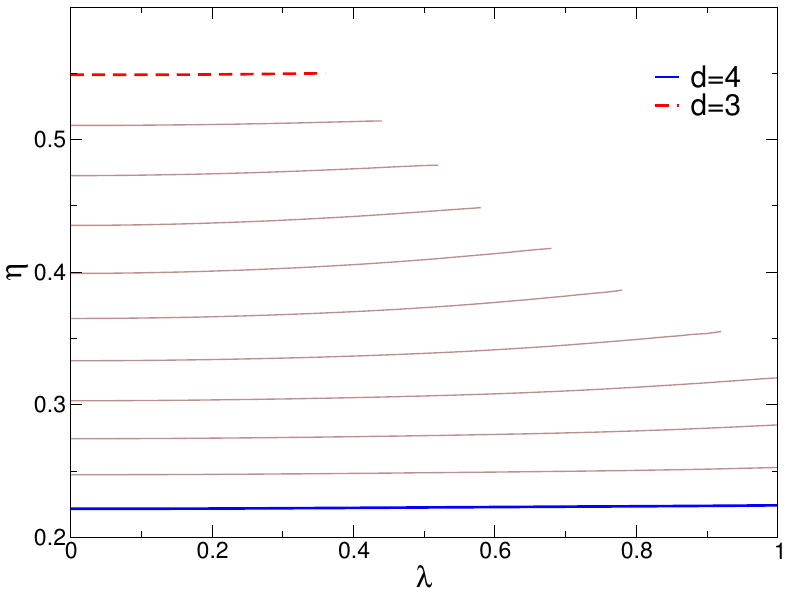}
\caption{Anomalous dimension $\eta(\lambda)$ obtained from the solution of the $\lambda$-modified FRG equations for several values of $d$ between 3 (top) and 4 (bottom).}
\label{Fig_B1}
\end{center}
\end{figure}

\begin{figure}
\begin{center}
\includegraphics[width=\linewidth]{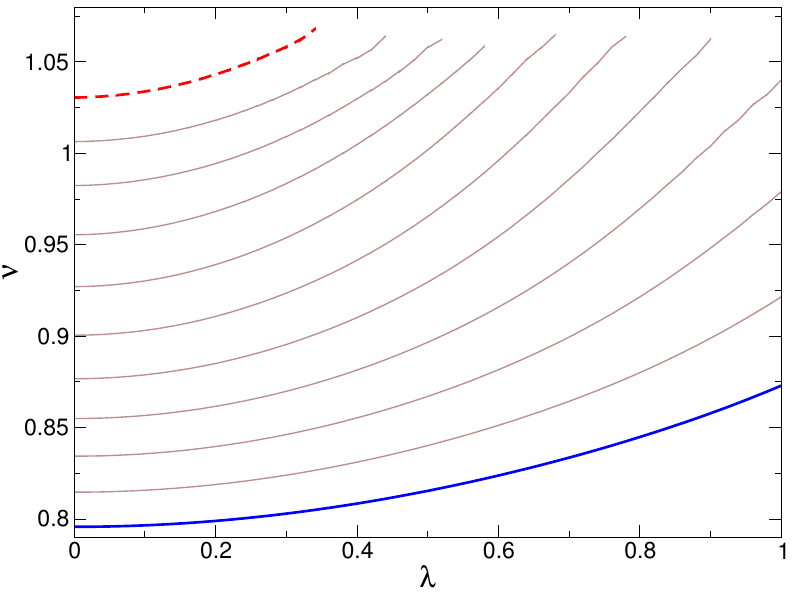}
\caption{Correlation length exponent $\nu(\lambda)$ obtained from the solution of the $\lambda$-modified FRG equations for several values of $d$ between 3 (top) and 4 (bottom).}
\label{Fig_B2}
\end{center}
\end{figure}

To obtain estimates for the hysteresis critical behavior when $d<3.5$ we have used a trick which amounts to introducing the same parameter $\lambda \in [0,1]$ in front of the anomalous contributions of all the RG beta functions: $\lambda=0$ corresponds to equilibrium and $\lambda=1$ to hysteresis. As $d$ is lowered below $3.5$ there is an increasing interval of $\lambda$ near $1$ for which we cannot solve these $\lambda$-modified FRG equations. However, we can solve the equations and obtain the critical exponents of the $\lambda$-modified theory for a whole range of $\lambda$ and $d$. We illustrate the results in Figs. \ref{Fig_B1} and \ref{Fig_B2} for the exponents $\eta$ and $\nu$. The dependences on $\lambda$ are smooth and have been fitted with polynomials. We can then extrapolate the data at each $d$ toward the value $\lambda=1$ to get an estimate for the hysteresis value. The extrapolated values are robust.

\subsection{Choice of the IR regulator}

For the computations, we have chosen an IR cutoff function (that enforces the decoupling of the fast and slow modes in the RG procedure) that has already been used in previous NP-FRG studies of the RFIM in equilibrium:\cite{tissier11}
\begin{equation}
r(y)=(a+by+c y^2)e^{-y}\,,
\end{equation}
where $y=q^2/k^2$. The parameters $a$, $b$, $c$ can be varied and, since the results such as the critical exponents (weakly) depend on the chosen values, the parameters can be in some sense optimized. To the least, what one requires is that the output, say the critical exponents, satisfies a property of ``minimum sensitivity" such that by varying the parameters around their optimum value, a minimal variation of the exponents results. For equilibrium it was shown that a good choice of parameters that provides robust results for a large range of dimensions is $a=1.7$, $b=0.81$, and $c=0.14$.\cite{tissier11} We have found here that varying the values around the latter ones (as well as  slightly changing the location of the fields where we constrain $z_k=1$ and $\delta_k=1$) has only a small influence on the exponents for both hysteresis and equilibrium criticality.


\begin{thebibliography}{35}
\bibitem{footnote_uppercritical}
Above the upper critical dimension, $d_{uc}=6$, the critical behavior for both equilibrium and hysteresis is of course controlled by the same Gaussian fixed point.
\bibitem{footnote_cuspy_eigenvalue}
Note that the equality of equilibrium and hysteresis critical exponents is also true for the exponent, denoted by $\lambda$ in Ref. [\onlinecite{tarjus13}], characterizing the least irrelevant cuspy perturbation. It is easy to show that the difference in the linearized RG equations brought by the anomalous contributions do not alter this eigenvalue.
\end{thebibliography}
\end{document}